%% file: Structure_review.tex
\documentclass[12pt,authoryear,a4paper]{elsarticle}

\usepackage{graphicx}

\usepackage{longtable}

\usepackage{xcolor}

\usepackage[colorlinks=true]{hyperref}
\hypersetup{pdftitle={Exploratory analysis of combined genome-wide SNP data}}
\hypersetup{pdfauthor={Blaise Li}}
\hypersetup{pdfsubject={Human population genetics}}
\hypersetup{pdfkeywords={Data combination, Graphical representation, Human populations, Negrito, Single nucleotide polymorphism}}
\hypersetup{unicode=true}
\hypersetup{breaklinks=true}

\PassOptionsToPackage{hyphens}{url}\usepackage{hyperref}

\usepackage{urllink}

\usepackage{bookmark}
\newcounter{labs}

\begin{document}
\sloppy

\begin{frontmatter}
\title{An exploratory analysis of combined genome-wide SNP data from several recent studies}

\author{Blaise Li}
\ead{blaise.li@normalesup.org}







\begin{abstract}
The usefulness of a `total-evidence' approach to human population genetics was
assessed through a clustering analysis of combined genome-wide SNP datasets.
The combination contained only 3146 SNPs.
Detailed examination of the results nonetheless enables the extraction of relevant
clues about the history of human populations, some pertaining to events as
ancient as the first migration out of Africa. The results are mostly coherent
with what is known from history, linguistics, and previous genetic analyses.
These promising results suggest that cross-studies data confrontation have the
potential to yield interesting new hypotheses about human population history.
\end{abstract}



\end{frontmatter}

\input{Main.tex}

\newpage
\input{Supplement.tex}

\end{document}

%% file: Main.tex
\section{Introduction}

Let this introduction begin with a disclaimer: I am not a population
geneticist, but a phylogeneticist who happens to be interested in human
population history. The results presented here should not be considered as
scientific claims about human population histories, but only as hypotheses that
might deserve further investigation.

In human population genetics, numerous papers have recently been published
using genome-wide SNP (Single Nucleotide Polymorphism) data for populations of
various places in the world. These papers often represent the data by means of
PCA (Principal Component Analysis) plots or clustering bar plots. The details
of such graphical representations suggest a variety of interesting hypotheses
concerning the relationships between populations. However, it is frustrating
to see the data scattered between different studies. Often, a study would
use data from other studies, but typically this would be limited to only a few
added populations. Would it not be possible and interesting to go further than
just adding the populations necessary to test some specific hypothesis? Do some
technical problems prevent the analyses of larger data combinations, involving
a wider range of populations. From my experience in phylogeny, I had been made
aware of the potential value of so-called `total-evidence' analyses, where
data combination helps extracting relevant information from noisy data.
Maybe something interesting could emerge from a total-evidence analysis
of these genome-wide SNP datasets.\\
I quickly noted that gathering the data from the published papers was more
difficult than expected. Data from human population genetics studies are not
as standardised as those used in phylogenetics. In particular, phylogenetic data
is usually stored in a centralised public database (NCBI Genbank) in a
standardised format. In human population genetics, it seems that each study has
its own policy regarding data availability, and its own way of storing it. In
the end, I could obtain the data from the \cite{HUGO_2009}, \cite{Reich_et_al_2009}
and \cite{Bryc_et_al_2010}, as well as those which are publicly available from
the HGDP \citep{Cann_et_al_2002,Li_et_al_2008b} and HapMap \citep{HapMap_2003} projects.\\
After struggling with the file formats and their different ways of coding the
genotypes, I could finally assemble the datasets into a single matrix, free from
the infamous A/T and G/C SNPs, and which seemed to produce reasonable results on
PCA plots (\emph{i.e.} a consistent placement of similar populations from
different datasets).

In the next section, I will describe and comment the results of clustering
analyses done with the program \texttt{frappe} \citep{Tang_et_al_2005}, in
growing number of clusters ($K$). For practical reasons, I decided to stop at
$K=16$. The clusters were becoming instable from one value of $K$ to the next.
This rendered the detailed examination of the results more difficult,
and unreasonably time consuming.

The figures were deposited as a file set on the FigShare repository:
\url{http://dx.doi.org/10.6084/m9.figshare.100442}.
The figures will be referenced using their individual \texttt{dx.doi.org} URL.


\section{Results}

\subsection{Graphical representation of the results}

For each clustering analysis, three kinds of bar plots were generated.\\
One series represents the profiles (proportions of each cluster) at the
individual level\footnote{\url{http://dx.doi.org/10.6084/m9.figshare.95764}}.
The list of clusters are reported below the graph, and for each cluster, the
population which has the highest average proportion of this cluster is
mentioned. The populations are grouped according to their region, their
language family and the alphabetical order of their names.\\
Another series represents the average profiles of the
populations\footnote{\url{http://dx.doi.org/10.6084/m9.figshare.95765}}. The
populations are grouped according to the geography, the language families, and
the profiles similarities.\\
The last series also represents the average profiles of the
populations, but there is one graph for each cluster, and for each
graph, the populations are ranked according to their proportion of the
corresponding cluster\footnote{\url{http://dx.doi.org/10.6084/m9.figshare.95784}}.

The colours were chosen based on language families and geography.
The language families are the first hierarchical levels of the classification
adopted by \cite{Lewis_et_al_2009}\footnote{\url{http://www.ethnologue.com/family_index.asp}}.

In the bar plots made at the individual level, an exception to the
grouping by geography and language family is made for the populations I labeled
`mixed', which I put in the end. Those populations were sampled in a region not
corresponding to their geographical origin or have a well-documented history of
admixture. It is of course somewhat arbitrary to decide which populations to
put in that separate category, as human population history is made of migration
and hybridization. For example, the Hakka and Minnan Chinese from Taiwan are
more recent inhabitants of the island than the Ami and Atayal Austronesians.
Their migration occurred roughly at the same time as the European and African
migrations to America. I could have labelled them as `mixed', since I have done
so with the `non-native' Americans. There are probably other similar cases; my
choices are inevitably biased by my perception of human population history.

Clusters are labelled by numbers. When comparing results obtained with
different values of $K$, to avoid ambiguities, I will often add a subscript to
the cluster number indicating the value of $K$ for which it was obtained.\\
Some clusters are well preserved from one value of $K$ to the next. In the
detailed description of the results, when such correspondences are not
discussed in the text, they are summarized in a table, using the
above-mentioned subscript notation.

The colour attributed to a cluster in the bar plots is determined by the
colour attributed to the population showing the highest proportion of that
cluster. This generally helps `tracking' a cluster across the different values of
$K$, except when populations with similar genetic profiles differ according to
their linguistic affiliations\refstepcounter{labs}\label{colours}. A small differential change in cluster
proportions between such populations may then lead to different colours being
attributed to `equivalent' clusters for different values of $K$. This is the case
when the European cluster is either most important in Basque or in Sardinians.

\subsection{Detailed results}

The detailed review of the results is available in annex (p.~\pageref{annex} and following).
It shows how clues about human population history can be extracted through close
examination. Readers interested in just having an idea of how this
information is extracted are invited to read the comments for the first values
of $K$ (up to $K=5$). More motivated readers may read the rest of the
description or even make their own examination of the figures.

\subsection{Summary of the results}

Average profiles of the populations at $K=2$:
\texttt{Frappe\_K2\_pops.pdf}\footnote{\url{http://dx.doi.org/10.6084/m9.figshare.188}}\\
At $K=2$, the separation in 2 clusters differentiates between an
`African' trend (cluster $1$) and an `East Asian' trend (cluster $2$).

Average profiles of the populations at $K=3$:
\texttt{Frappe\_K3\_pops.pdf}\footnote{\url{http://dx.doi.org/10.6084/m9.figshare.95713}}\\
At $K=3$, the 3 trends are `African' (cluster $1$), `European' (cluster $2$) and
`East Asian' (cluster $3$).

Average profiles of the populations at $K=4$:
\texttt{Frappe\_K4\_pops.pdf}\footnote{\url{http://dx.doi.org/10.6084/m9.figshare.189}}\\
At $K=4$, an `American' cluster (number $4$) is added to the three
previous ones: `African' (number $1$), `European' (number $2$) and `East Asian' (number
$3$).

Average profiles of the populations at $K=5$:
\texttt{Frappe\_K5\_pops.pdf}\footnote{\url{http://dx.doi.org/10.6084/m9.figshare.190}}\\
At $K=5$, there is one cluster for each continent:
\begin{itemize}
\item cluster $1$, the `African' cluster (more specifically, `Sub-Saharan');
\item cluster $2$, the `European' cluster;
\item cluster $3$, the `Asian' cluster (more specifically, `East Asian');
\item cluster $4$, the `Oceanian' cluster;
\item cluster $5$, the `American' cluster.
\end{itemize}
This result is comparable to what has been already obtained with the HGDP
sample \citep{Cann_et_al_2002}.

Average profiles of the populations at $K=6$:
\texttt{Frappe\_K6\_pops.pdf}\footnote{\url{http://dx.doi.org/10.6084/m9.figshare.191}}\\
At $K=6$, the `East Asian' cluster $3_5$ is split into a `northern' component (cluster
$3_6$) and a `southern' component (cluster $4_6$).

Average profiles of the populations at $K=7$:
\texttt{Frappe\_K7\_pops.pdf}\footnote{\url{http://dx.doi.org/10.6084/m9.figshare.192}}\\
At $K=7$, the new cluster that appears, number $2_7$, having its highest
frequencies in Dravidian populations, and more generally in India and Pakistan,
represents a `South Asian' tendency. This cluster seems to principally replace
parts of the `European' ($2_6$) and `Oceanian' ($5_6$) clusters.

Average profiles of the populations at $K=8$:
\texttt{Frappe\_K8\_pops.pdf}\footnote{\url{http://dx.doi.org/10.6084/m9.figshare.193}}\\
At $K=8$, a `non-Niger-Congo' cluster ($2_8$) replaces part of the previous `African'
($1_7$) and `European' ($3_7$) clusters.

Average profiles of the populations at $K=9$:
\texttt{Frappe\_K9\_pops.pdf}\footnote{\url{http://dx.doi.org/10.6084/m9.figshare.194}}\\
At $K=9$, the `southern East Asian' cluster which was dominant in Mlabri
($6_8$) is decomposed in two clusters ($6_9$ and $7_9$). There are now 3
`East Asian' clusters:
\begin{itemize}
\item Cluster $4_9$ is more present in Altaic, Korean and Japanese populations.
\item Cluster $6_9$ is more present in Austronesian populations.
\item Cluster $7_9$ is typical of Malaysian Negritos.
\end{itemize}

Average profiles of the populations at $K=10$:
\texttt{Frappe\_K10\_pops.pdf}\footnote{\url{http://dx.doi.org/10.6084/m9.figshare.195}}\\
At $K=10$, Mlabri have their profile exclusively composed of cluster
$7_{10}$, which partly substitutes the `Austronesian' and `southern East Asian'
clusters $6_9$ (then $6_{10}$) and $7_9$ (then $8_{10}$).

Average profiles of the populations at $K=11$:
\texttt{Frappe\_K11\_pops.pdf}\footnote{\url{http://dx.doi.org/10.6084/m9.figshare.196}}\\
At $K=11$, the `African' trend is now divided in 3 clusters. A new
`Khoisan-Pygmy' cluster ($2_{11}$) is added to the previously identified `general
Sub-Saharan' and `East African-West Asian' cluster.

Average profiles of the populations at $K=12$:
\texttt{Frappe\_K12\_pops.pdf}\footnote{\url{http://dx.doi.org/10.6084/m9.figshare.197}}\\
At $K=12$, the `Khoisan-Pygmy' cluster disappears, and a rearrangement
of the `East Asian' clusters occurs:
\begin{itemize}
\item There are 2 `Austronesian' clusters ($6_{12}$ and $7_{12}$), one
of which ($6_{12}$) is in fact more specific to the non-Filipino populations of
the Philippines. Cluster $7_{12}$ has a reinforced `Austronesian' character.
\item A `continental South-East Asian' cluster appears.
\item The `northern East Asian' cluster $4$ acquires a more `maritime' flavour.
\item The `Mlabri-specific' and `Malaysian Negrito-specific' clusters are
maintained.
\end{itemize}

Average profiles of the populations at $K=13$:
\texttt{Frappe\_K13\_pops.pdf}\footnote{\url{http://dx.doi.org/10.6084/m9.figshare.198}}\\
At $K=13$, there are several important changes:
\begin{itemize}
\item The `Khoisan-Pygmy' cluster observed at $K=11$ reappears.
\item A new `Middle Eastern' cluster ($4_{13}$) appears.
\item The cluster specific to the Negritos from the Philippines ($6_{12}$) disappears.
\end{itemize}

Average profiles of the populations at $K=14$:
\texttt{Frappe\_K14\_pops.pdf}\footnote{\url{http://dx.doi.org/10.6084/m9.figshare.199}}\\
At $K=14$, the `Middle Eastern' cluster disappears, but the `Khoisan-Pygmy' cluster
is still there. The Asian clusters are highly reorganized:
\begin{itemize}
\item There are two `Austronesian' clusters. Cluster $7_{14}$ is dominant
in Borneo, Java and the Malaysian peninsula and cluster $8_{14}$ is dominant in
the Philippines.
\item There is a `southern East Asian' cluster ($11_{14}$) predominant
in Hmong-Mien and Sino-Tibetan populations.
\item There is a cluster specific to the Andamanese and Negritos from
the Philippines ($12_{14}$).
\item The `Indian' ($4_{14}$), `northern East Asian' ($5_{14}$),
`Mlabri-specific' ($9_{14}$), and `Malaysian Negrito' ($10_{14}$) clusters can
still be identified.
\end{itemize}

Average profiles of the populations at $K=15$:
\texttt{Frappe\_K15\_pops.pdf}\footnote{\url{http://dx.doi.org/10.6084/m9.figshare.200}}\\
At $K=15$, a `Middle Eastern' cluster is present, as was the case at $K=13$.
The other clusters correspond to those present at $K=14$.

Average profiles of the populations at $K=16$:
\texttt{Frappe\_K16\_pops.pdf}\footnote{\url{http://dx.doi.org/10.6084/m9.figshare.201}}\\
At $K=16$, the cluster specific to the Andamanese populations again
disappears. The `Austronesian' clusters are reorganized, with the
appearance of a cluster specific to the non-Filipino populations of the
Philippines ($10_{16}$), as was the case at $K=12$. The `American'
cluster is now separated in two:
\begin{itemize}
\item Cluster $15_{16}$ is more present in North America, and is almost absent in
the Tupi-speaking populations from the Amazon forest (Surui and Karitiana).
\item Cluster $16_{16}$ is highly dominant in the Tupi, but is also present in the
other American populations.
\end{itemize}

\section{Discussion}\refstepcounter{labs}\label{discussion}

In this section, I will sometimes use distance trees to compare the profiles of
the populations. I will call such trees `profile trees' (see Materials and
Methods, p.~\pageref{profile_trees}). It should be noted that these do not aim to
represent historical relationships between populations, but only similarities
between their clustering profiles\footnote{The profile trees will contain
clusters of clustering profiles, but it should be clear from the context what
type of cluster a sentence is about.}. The similarities between clustering
profiles are however likely to partially reflect historical relationships, and
can therefore be used as an exploratory tool to investigate such relationships.

\subsection{Correlations with geography}

Not surprisingly, like in the original studies of the individual datasets,
the compositions of the profiles are mainly correlated with geography. For
example, in the profile tree for
$K=16$\footnote{\url{%http://hdl.handle.net/10779/K16_allpops.profile_tree
http://dx.doi.org/10.6084/m9.figshare.216}},
one can clearly see a cluster containing the populations of Sub-Saharan Africa,
one containing the populations of North Africa, Middle East, Europe and
Caucasus and one containing almost all populations of Pakistan and India (the
exceptions being the Tibeto-Burmese-speaking populations, the Himalayan Pahari
and the Hazara, which are closer to the cluster containing the populations of
Central, North, and East Asia, the Siddi, which are closer to the Sub-Saharan
cluster, and the reciprocal exception are the Indians from Singapore, which
cluster with the populations of India).\\
Within the main clusters, other smaller clusters can be found that reflect
geography. For example, the populations of the Lesser Sunda Islands cluster
with Papuans and Melanesians.\\
Geographic structure may also be evidenced within a subset of the populations.
For example, in profile trees using populations from west and south
Eurasia\footnote{The trees include the populations of Europe, Caucasus, Middle
East, Pakistan (except Hazara), and mainland India (except Pahari and
Tibeto-Burmese).}, for most values of $K$, the populations are disposed along
the tree in an order that correlates quite well with a west $\leftrightarrow$
east direction: Europe, Middle East, Caucasus, Pakistan, Kashmir, and the rest
of India\footnote{See for example
\url{%http://hdl.handle.net/10779/K8_west_and_south_Eurasia.profile_tree
http://dx.doi.org/10.6084/m9.figshare.223}.}.
The differentiation between Pakistan, Kashmir, and the rest of India parallels
the north-Indian / south-Indian opposition evidenced in
\cite{Reich_et_al_2009}, but with less details within India. This lack of detail
could be due to a much smaller number of SNPs, and also to a less conservative
way of selecting populations.


\subsection{A note on Negritos and the southern route}


As early as $K=3$, the presence of the `African' cluster in some populations of
South and South-East Asia and Oceania was noticed and interpreted as a possible
trace of an old genetic background dating back to early waves of migration out
of Africa (see annex, p.~\pageref{suppl:early}). Among these
populations, Papuans, Melanesians, Andamanese and Negritos from the Philippines
and the Malaysian peninsula share the particularity of having a morphology in
some points similar to the populations of Africa\footnote{This morphological
particularity led the Spanish to use the term `Negrito' for some populations of
the Philippines. This term is also used for the hunter-gatherer populations of
the Malaysian peninsula, and sometimes also for the Andamanese populations.}.
This is often interpreted as adaptive convergence, because, from the genetic
point of view, these populations have no striking similarities. As we shall
see, a closer examination of the genetic data reveals that the overall genetic
disparity of these populations hides a few intriguing similarities.

The interpretation of the presence of the `African' cluster in Oceanian
populations and ANLS (Andaman, Negrito, Lesser Sunda) as an `early wave'
signature is reinforced when one considers what happens when the `Oceanian'
cluster appears, at $K=5$. The `African' cluster not only decreases in
Papuans, Melanesians and in the populations of the geographically close Lesser
Sunda Islands, but also in the more remote Andamanese and Negritos from the
Malaysian peninsula and from the Philippines, while the decrease is much lower
in populations of recent African ancestry (see annex, p.~\pageref{suppl:out-of-Africa}).
This sharing of profile co-variation by scattered populations is best explained
by a shared ancient genetic background, dating to a time when the sea level
was lower, than by more recent population migrations. Indeed, contrary to other
populations of maritime South-East Asia that are well known for their mastery of
navigation, Andamanese and Negritos from the Malaysian peninsula and from the
Philippines are land-bound hunter-gatherers. But their lifestyle could of course
have changed: The case of Mlabri suggests that a `reversion' to a hunter-gatherer
lifestyle may happen \citep{Oota_et_al_2005}.

At $K=11$ another interesting observation arises from the appearance of a
cluster dominant in San and Pygmies. First, this shows that Khoisan and
Pygmies, all traditionally hunter-gatherers, share not only a mode of
subsistence, but also some genetic characteristics. Since they are scattered in
various places of Sub-Saharan Africa, this could be interpreted as shared
ancestry, dating before the spread of the Bantu populations. A less visible
consequence of the appearance of the `Khoisan-Pygmy' cluster is a
differential split of the `African ancestry' of populations outside Africa into
the different `African' components. The portion of putative African ancestry which
is represented by the `Khoisan-Pygmy' cluster is higher in ANLS than in the populations
of recent African ancestry (see annex, p.~\pageref{suppl:hunter-gatherers}).

It should be also noted that when the `Khoisan-Pygmy' cluster disappears at
$K=12$, the `Austronesian' cluster is split in two, one of the resulting
clusters ($6_{12}$) having its highest proportion in the Negritos from the
Philippines Mamanwa, Ati, Ayta and Agta\footnote{See
\url{http://dx.doi.org/10.6084/m9.figshare.301}.}. This cluster disappears at
$K=13$, while the `Khoisan-Pygmy' cluster reappears. These
switches\refstepcounter{labs}\label{pygmy-negrito-balance} between the presence of one or the other
cluster suggests that some aspect of the genetic composition of the Negritos
from the Philippines can be either accounted for by the presence of a
`Khoisan-Pygmy' cluster or by a more specific cluster.

The particularity of the African ancestry of ANLS populations can also be
evidenced by PCA (Principal Component Analysis). The \texttt{smartpca} program of the
EIGENSOFT package \citep{Patterson_et_al_2006} allows the determination of the
principal components using only a subset of the analyzed populations (option
\texttt{-w}). I used a selection of Sub-Saharan populations (including Pygmies
and San, but excluding the atypical Maasai, Luhya and Fulani) to determine the
principal components, and then generated the PCA plot of the populations of
interest using the first two principal components. The first component
differentiates between a `Khoisan-Pygmy' side and a `general Sub-Saharan' side.
The second principal component reveals the disparity between San, Biaka and Mbuti.
Plotting each individual does not allow to see a clear trend, but representing
the populations using the averages of the coordinates of their individuals
does\footnote{See \url{http://dx.doi.org/10.6084/m9.figshare.23398}.}.\\
The populations with recent known or possible African ancestry tend to be situated
on the `general Sub-Saharan' side, while ANLS populations and Papuans (who could
also bear the genetic traces of the first migrants out of Africa) occupy a more
intermediate position, as do the south-eastern Bantu populations (who have
received genetic input from Khoisan populations). The principal component that
differentiates between Khoisan and Pygmy, on one side, and other Sub-Saharan
populations on the other side, also differentiates between ANLS and Papuans on
one side, and populations of recent African ancestry on the other side.

These observations suggest that (if the `early wave' origin of the African
component detected in ANLS is accepted) the early out-of-Africa migrants did
hold a share of the African genetic diversity more similar to that retained by
Khoisan and Pygmies than that retained by other African populations (see
annex, p.~\pageref{suppl:hunter-gatherers}). Another fact that supports this hypothesis is that
the morphological characteristics shared by some ANLS populations with Khoisan
and Pygmies are not only general features of African populations such as skin colour
and hair type, but also more specific characteristics, like short stature. Quite
interestingly, Onge and Pygmy women are even subject to steatopygia, an uncommon
physical feature for which Khoisan are well known. It would be interesting to test
whether these shared characteristics could be inherited from a common ancestor, rather
than simply be adaptive convergences.

\subsection{Austronesian affinities}

The PASNP data for Asian populations \citep{HUGO_2009} used in the present work
concern a large number of populations and a relatively smaller number of SNPs
than the other datasets. Since the dataset combination consisted in an union of
the populations and in an intersection of the SNPs, the assembled dataset
probably carries more detailed information for Asian populations than for the
other parts of the world. In particular, this permitted marked distinctions
between Austronesians. Among these populations, for high values of $K$, the
following groups can be distinguished\footnote{See for example
\url{%http://hdl.handle.net/10779/K15_Austronesians.profile_tree
http://dx.doi.org/10.6084/m9.figshare.244}.}:
\begin{itemize}
\item populations of the Lesser Sunda Islands;
\item Iraya and Negritos from the Philippines;
\item Mentawai, Toraja, Manobo, Filipinos and Taiwanese (the latter two being
more often grouped together);
\item populations of the Malaysian peninsula, Sumatra (except Mentawai), Java
and Borneo, with the following subgroups:
\begin{itemize}
\item Batak and Malays;
\item Temuans and populations of Java and Borneo.
\end{itemize}
\end{itemize}

Below $K=12$, the cluster containing the populations of the Lesser Sunda
Islands is included in the cluster containing the Negritos from the
Philippines, and Iraya tend to form a more distant branch\footnote{See for
example \url{%http://hdl.handle.net/10779/K12_Austronesians.profile_tree
http://dx.doi.org/10.6084/m9.figshare.241}.}.
Below $K=7$, the clusters tend to disaggregate\footnote{See for example
\url{%http://hdl.handle.net/10779/K5_Austronesians.profile_tree
http://dx.doi.org/10.6084/m9.figshare.235}.}.

On profile trees including Tai-Kadai and Austronesian populations, Tai-Kadai
tend to cluster with Taiwanese and Filipinos. This is approximately the case
from $K=2$ to $K=5$\footnote{See for example
\url{%http://hdl.handle.net/10779/K4_Austro-Tai.profile_tree
http://dx.doi.org/10.6084/m9.figshare.248}.}, and exact for
$K=6$ to $K=11$ and at $K=13$\footnote{See for example
\url{%http://hdl.handle.net/10779/K6_Austro-Tai.profile_tree
http://dx.doi.org/10.6084/m9.figshare.250}.}, but with a
growing branch length for the Tai-Kadai sub-group as $K$ increases\footnote{See
for example \url{%http://hdl.handle.net/10779/K11_Austro-Tai.profile_tree
http://dx.doi.org/10.6084/m9.figshare.255}.}. At
$K=12$, $K=14$, $K=15$ and $K=16$, Tai-Kadai form a separate
cluster\footnote{See for example
\url{%http://hdl.handle.net/10779/K14_Austro-Tai.profile_tree
http://dx.doi.org/10.6084/m9.figshare.258}.}.\\
If Tai-Kadai have a part of Austronesian ancestry, the profile similarities
between Tai-Kadai, Taiwanese and Filipinos suggest that the Austronesian
ancestors of Tai-Kadai populations were probably an early offshoot of the
Austronesian dispersal (hypothesized to have started from Taiwan).
This is compatible with the linguistic evidence detailed in \cite{Sagart_2004}
(see also annex, p.~\pageref{suppl:Austro-Tai}).
However, in the profile trees including all populations, this relationship
between Tai-Kadai and `basal' Austronesians is obscured by the fact that,
depending on the value of $K$, Tai-Kadai sometimes cluster with Chinese and
Hmong-Mien populations\footnote{See for example
\url{%http://hdl.handle.net/10779/K12_allpops.profile_tree
http://dx.doi.org/10.6084/m9.figshare.212}.}. Moreover, Mon-Khmer and
JKL (Jinuo, Karen, Lahu) populations sometimes also cluster with
Austronesians\footnote{See for example
\url{%http://hdl.handle.net/10779/K9_allpops.profile_tree
http://dx.doi.org/10.6084/m9.figshare.209}.}. For high values of
$K$ the non-Mlabri and non-Negrito Mon-Khmer populations tend to cluster with
JKL, Temuans and the populations of Java and Borneo\footnote{See for example
\url{%http://hdl.handle.net/10779/K15_allpops.profile_tree
http://dx.doi.org/10.6084/m9.figshare.215}.}.

One may regret the absence of Polynesians (easternmost Austronesians), Malagasy
(Austronesians who migrated to the west of the Indian Ocean) and Cham (see the
discussion concerning the presence of cluster $7_{12}$ in Cambodians,
p.~\pageref{suppl:Cham} of the annex) populations in the dataset.
This would have offered an even better coverage of the diversity of the
Austronesian populations.

\subsection{Trans-linguistic affinities}

A few trans-linguistic clusters repeatedly appear in the profile trees. Besides
the above-mentioned grouping of the populations of the Lesser Sunda Islands
with Melanesians and Papuans, one should notice the grouping of the
Indo-Iranian Hazara with the Altaic Uyghur. This constitutes a strong evidence
for attributing Hazara an origin in Central Asia. Another atypical Indo-Iranian
population are the Pahari, which group with Tibeto-Burmese Spiti. Their profile
similarities probably reflect genetic exchanges between Tibeto-Burmese and
Indo-Iranian populations in the Himalayan region (see also annex,
p.~\pageref{suppl:Himalaya} and p.~\pageref{suppl:Himalaya-2}). A third
trans-linguistic grouping involving an Indo-Aryan population is that of
Sahariya with Munda. It appears repeatedly, and in some trees, these
populations also group with Andamanese. It is difficult to tell whether this
might be due to some shared ancestry or if this is only an effect of convergent
hybridization events between similar Asian genetic stocks. Indeed, the grouping
of Fulani with African Americans (and sometimes also with the Maasai) suggests
that obviously different histories may produce similarities in the profiles.


\subsection{Contrasts within a linguistic family}

Differences internal to a linguistic group are also revealed by the comparison
of profiles. Different groups of Austronesian populations have been
discussed earlier. Other conspicuous cases of `intra-linguistic' differences
can be observed. An interesting example is offered by the Sino-Tibetan family.
On profile trees including Sino-Tibetan, Hmong-Mien and Tai-Kadai populations,
besides the long branch of the Himalayan Spiti, a striking fact is the
particularity of the Tibeto-Burmese populations from the Burmese border (JKL).
For most values of $K$, the profile tree is `linear', with the populations in
the following sequence: Spiti, Tibeto-Burmese of east India (Nysha and Aonaga),
Tibeto-Burmese of inner south China (Naxi and Yizu), northern Chinese, Tujia,
southern Chinese and She, other Hmong-Mien, eastern Tai-Kadai, western
Tai-Kadai, JKL\footnote{See for example
\url{%http://hdl.handle.net/10779/K14_Sino-Hmong-Mien-Tai.profile_tree
http://dx.doi.org/10.6084/m9.figshare.273}.}. The
JKL have thus profiles quite distinct from those of the other
Tibeto-Burmese populations, and in particular distinct from Naxi and Aonaga,
which were not sampled very far from the Burmese border, but at more northern
locations. Karen, Jinuo and Spiti were listed among the `linguistic outliers'
in the original publication of the data \citep[p.~1543]{HUGO_2009}.
To be also noted on these profile trees is the difference between
the She (which have profiles similar to the neighbouring southern Chinese) and
the other Hmong-Mien populations (whose profiles are intermediate between
southern Chinese and Tai-Kadai profiles).

Less conspicuous intra-linguistic differences can also be detected on the profile trees.
For low values of $K$, Druze appear to have a profile more similar to European
populations than to Palestinians and Bedouins sampled in the same region\footnote{See for example
\url{%http://hdl.handle.net/10779/K4_allpops.profile_tree
http://dx.doi.org/10.6084/m9.figshare.204}.}.
The Druze community has its origins at the beginning of the 11th century in the
multi-ethnic Fatimid empire. Among its founders are people of Persian and Turk
origins, and some famous Druze family names suggest Kurd (Jumblatt) or Turk
(Arslan) origins. It may thus be hypothesized that a non-Arab genetic
contribution explains the small differences observed between the
profiles of Druze and those of the two other populations from Middle East.


\subsection{Profiles co-variation patterns}

I will suggest here another manner of using the clustering analyses as an
exploratory tool. If the clustering profiles of two population `react' in the
same manner when the clusters are reorganised (that is, when $K$ changes), this
may be a sign that these populations share a portion of genetic ancestry
inherited from a common population. Therefore, besides considering the direct
similarities between profiles, it may be useful to also pay attention to
recurrent co-variation patterns\footnote{One could even devise some ways of
automatically proposing a correspondence between clusters for different values
of $K$, use this to compute vectors of `derivatives' of the ancestry
profiles for the populations, and build distance trees between these vectors,
in order to facilitate the detection of such co-variation patterns.}.

For example, some co-variations are observed between the profiles of the
populations of Japan, Taiwan and the Philippines:
\begin{itemize}
\item When comparing $K=12$ with $K=10$, a rank decrease for the `Indian'
cluster $3$ was observed in the Philippines, Taiwan and Japan, and a rank
increase occurred for Filipinos and Taiwanese Austronesians for the `northern
East Asian' cluster $4$, while the contrast between the populations of Japan and
the other populations of northern East Asia was reinforced (see
annex, p.~\pageref{suppl:Japan-Austronesian-1}).
\item When comparing the situations at $K=11$ and $K=13$ increases in the
`Mlabri-specific' and `Malaysian Negrito' clusters were observed in the
Philippines, Taiwan and Japan (see annex, p.~\pageref{suppl:Japan-Austronesian-2}).
\item When comparing the situations at $K=12$ and $K=14$, an increase in the
`southern East Asian' cluster was observed in Taiwan, Japan and the
Philippines (see annex, p.~\pageref{suppl:Japan-Austronesian-3}).
\end{itemize}

It can be noticed in this respect that the Austronesian populations that have
the highest proportion of the northern `East Asian' cluster (which is
dominant in Japan) are Filipinos and Taiwanese Austronesians, for all values of
$K$ for which this cluster exists (that is, from $K=6$ and above).

A possible explanation for these observations could be the maritime activity
that occurred in historical times in the region, for instance through Ryukyuan
traders. This would have eased the sharing of genetic characteristics between
the populations of Taiwan, Japan and the Philippines. More recent events can
also be invoked, such as the colonization of Taiwan by the Japanese empire or
Japanese migrations to the Philippines during the first half of the 20th
century. 

Another example is that some co-variations are observed between the profiles
of Okinawans and of the populations of the Andaman islands:
\begin{itemize}
\item When comparing the situations at $K=11$ and $K=13$, a simultaneous
decrease was observed in the `Oceanian' cluster for Okinawans and Onge
(see annex, p.~\pageref{suppl:Okinawa-Andaman-1}).
\item The `Oceanian' cluster decreased in Andamanese populations at
$K=14$, when the cluster specific to Andamanese populations appeared
($12_{14}$), and a strong rank decrease was then observed in that cluster for
Okinawans (see annex, p.~\pageref{suppl:Okinawa-Andaman-2}).
\end{itemize}

These correlations could make sense in the light of the fact that both
Andamanese and Okinawans have been reported to have a high proportion of Y
chromosome haplogroup D \citep[see][p.~51 and p.~55]{Hammer_et_al_2006}.
This would reflect an ancient genetic background shared by these two
populations. It could be interesting in this respect to add Ainu samples
to the dataset, in order to have a better picture of the ancient genetic
landscape of Japan.

Yet another example of co-variation pattern is the already mentioned switches
between the presence of a `Khoisan-Pygmy' cluster and one specific to the
Negritos from the Philippines (see p.~\pageref{pygmy-negrito-balance}). These
switches concur in suggesting to investigate the possibility that Negrito
populations could share some ancient genetic background with Pygmies and
Khoisan populations.

\section{Conclusions}

When the analyses were performed, the data available from the PASNP consortium
did only contain autosomal SNPs. The combined dataset does therefore not
contain SNPs located in the Y or mitochondrial chromosomes. The results
obtained here are thus complementary to what can be inferred from the studies
of Y or mtDNA haplogroups.

If the clusters are to be interpreted as ancestry classes, low values of $K$
might reflect inheritance from older ancestral populations than high values of
$K$. Although more accurate for describing similarities between extant
populations, bar plots made with high values of $K$ would then be less likely to
reflect ancient historical events. By focusing only on one value of $K$, or on
a narrow range, one might miss some clues about population history. I would
therefore suggest that a wide range of values of $K$ be considered
when clustering analyses are used as an exploratory tool.

Despite the small number of SNPs in the combined dataset, the clustering bar
plots seem to convey a significant amount of relevant information about human
population history\footnote{Preliminary analyses using one more source in the
combination \citep[the data from][]{Xing_et_al_2010} indicate that similar
clustering patterns are obtained using only 1656 SNPs. See
\url{http://dx.doi.org/10.6084/m9.figshare.89584}}.
Therefore, the practice consisting in combining data at a large geographical
scale seems promising and should be tried with an even more diverse population
sampling. This `taxonomical total-evidence' approach (I borrow here vocabulary
from phylogenetics) would be facilitated if the data were stored in a central
repository, under a standardised format, and could be more powerful with a better
SNP overlap between studies.

Although this work probably does not bring many new results in human population
history, I enjoyed the experience and hope that my remarks from outside can be
useful to the community of human population genetics.



\section{Materials and Methods}\refstepcounter{labs}\label{MM}

\subsection{Data preparation}

The SNP data were obtained from the following sources:
\begin{itemize}
\item `HGDP' \citep{Cann_et_al_2002,Li_et_al_2008b}: the Stanford University HGDP-CEPH SNP
genotyping data, supplement 1 (1043 samples);
\item `HapMap' \citep{HapMap_2003}: draft release 2 for the genome-wide SNP
genotyping of the phase 3 samples (1184 samples);
\item `Asia' \citep{HUGO_2009}: the PASNP consortium genotype data (1928 samples,
only the autosomal SNPs were included in the present study);
\item `India' \citep{Reich_et_al_2009}: SNP data for various populations of India,
including populations from the Andaman Islands (132 samples);
\item `Africa' \citep{Bryc_et_al_2010}: SNP data for various populations of Africa
(370 samples).
\end{itemize}

According to \url{http://www.cephb.fr/common/RosenbergPreprint.pdf}, the HGDP
samples include related individuals and 13 duplicates, one of which is labelled
both as a Hazara and as a Pathan individual. The duplicates were apparently
already suppressed from the downloaded dataset, and the bi-labelled individual
completely removed. I had to remove the mis-labelled Biaka Pygmy and Japanese
individuals reported in that same document.\\
Some of the HapMap samples are grouped in (mother, father, child) triplets. For
such samples, the child was removed.

The data for all remaining samples were combined using \texttt{python}
(\url{http://www.python.org/}) scripts, keeping only the SNPs that were present
in the five datasets. The format of the source data differed, and it was not
always clear how SNP states between 2 datasets compared. PCA analyses using the
\texttt{smartpca} program \citep{Patterson_et_al_2006} did not show obvious
inconsistencies when comparing geographically close populations from different
datasets. The resulting combined dataset consists in the genotypes of 4025
individuals at 3146 SNPs. The distribution of the SNPs is summarized in the
following table:\\
\\
{\small
\begin{tabular}{|c|c|c|c|c|c|c|c|c|c|c|c|}
\hline
chromosome &
1  &
2  &
3  &
4  &
5  &
6  &
7  &
8  &
9  &
10 &
11 \\\hline
\# of SNPs &
262 &
264 &
203 &
222 &
241 &
209 &
175 &
166 &
132 &
178 &
166 \\
\hline\hline
chromosome &
12 &
13 &
14 &
15 &
16 &
17 &
18 &
19 &
20 &
11 &
22 \\\hline
\# of SNPs &
160 &
146 &
115 &
99  &
71  &
74  &
100 &
22  &
76  &
45  &
20 \\
\hline
\end{tabular}}

Some populations are sampled in more than one dataset, under different names
(for example Uyghur in \cite{HUGO_2009} and Uygur in \cite{Li_et_al_2008b}). I
kept the original names. The populations are thus distinguished in the admixture
graphs, but I used only one spelling in the present text. The two samples
did not need to be distinguished in the comments, given the high similarity
of their clustering profiles.

The following table gives the list of the sampled populations, with the
associated linguistic information:
{\small
\begin{longtable}{|l|l|l|}
\hline
Population & Language group & Language sub-group \\
\hline
\endhead
\hline
\endfoot
\textcolor{green}{Adygei} & North-Caucasian & West-Caucasian \\
\textcolor{teal!75!green}{African American} & Indo-European & Germanic \\
\textcolor{blue!66!orange}{Agta} & Austronesian & Malayo-Polynesian \\
\textcolor{blue!66!orange}{Alorese} & Austronesian & Malayo-Polynesian \\
\textcolor{blue!66!orange}{Ami} & Austronesian & East-Formosan \\
\textcolor{yellow!75!lime!90!black}{Aonaga} & Sino-Tibetan & Tibeto-Burman \\
\textcolor{blue!66!orange}{Atayal} & Austronesian & Atayalic \\
\textcolor{blue!66!orange}{Ati} & Austronesian & Malayo-Polynesian \\
\textcolor{blue!66!orange}{Ayta} & Austronesian & Malayo-Polynesian \\
\textcolor{teal!75!green}{Balochi} & Indo-European & Indo-Iranian \\
\textcolor{red}{Bamoun} & Niger-Congo & Atlantic-Congo \\
\textcolor{red}{Bantu NE} & Niger-Congo & Atlantic-Congo \\
\textcolor{red}{Bantu SE Pedi} & Niger-Congo & Atlantic-Congo \\
\textcolor{red}{Bantu SE Sotho} & Niger-Congo & Atlantic-Congo \\
\textcolor{red}{Bantu SE Tswana} & Niger-Congo & Atlantic-Congo \\
\textcolor{red}{Bantu SE Zulu} & Niger-Congo & Atlantic-Congo \\
\textcolor{red}{Bantu SW Herero} & Niger-Congo & Atlantic-Congo \\
\textcolor{red}{Bantu SW Ovambo} & Niger-Congo & Atlantic-Congo \\
\textcolor{blue!66!orange}{Batak Karo} & Austronesian & Malayo-Polynesian \\
\textcolor{blue!66!orange}{Batak Toba} & Austronesian & Malayo-Polynesian \\
\textcolor{purple!75!lime}{Bedouin} & Afro-Asiatic & Semitic \\
\textcolor{teal!75!green}{Bengali} & Indo-European & Indo-Iranian \\
\textcolor{teal!75!green}{Bhil} & Indo-European & Indo-Iranian \\
\textcolor{teal!75!green}{Bhili} & Indo-European & Indo-Iranian \\
\textcolor{red}{Biaka Pygmies} & Niger-Congo & Atlantic-Congo \\
\textcolor{blue!66!orange}{Bidayuh Jagoi} & Austronesian & Malayo-Polynesian \\
\textcolor{lime!75!cyan!90!black}{Brahui} & Dravidian & Northern-Dravidian \\
\textcolor{red}{Brong} & Niger-Congo & Atlantic-Congo \\
\textcolor{purple}{Bulala} & Nilo-Saharan & Central-Sudanic \\
\textcolor{lime!90!black}{Burusho} & Burushaski & Burushaski \\
\textcolor{orange!75!blue}{Cambodians} & Austro-Asiatic & Mon-Khmer \\
\textcolor{lime!75!cyan!90!black}{Chenchu} & Dravidian & South-Central-Dravidian \\
\textcolor{yellow!75!lime!90!black}{Chinese Denver} & Sino-Tibetan & Chinese \\
\textcolor{yellow!75!lime!90!black}{Chinese Hakka} & Sino-Tibetan & Chinese \\
\textcolor{yellow!75!lime!90!black}{Chinese Minnan} & Sino-Tibetan & Chinese \\
\textcolor{violet}{Colombians} & Arawakan & Maipuran \\
\textcolor{orange}{Dai} & Tai-Kadai & Kam-Tai \\
\textcolor{olive!66!teal}{Daur} & Altaic & Mongolic \\
\textcolor{blue!66!orange}{Dayak} & Austronesian & Malayo-Polynesian \\
\textcolor{purple!75!lime}{Druze} & Afro-Asiatic & Semitic \\
\textcolor{teal!75!green}{European Utah} & Indo-European & Germanic \\
\textcolor{red}{Fang} & Niger-Congo & Atlantic-Congo \\
\textcolor{blue!66!orange}{Filipino Ilocano} & Austronesian & Malayo-Polynesian \\
\textcolor{blue!66!orange}{Filipino Tagalog} & Austronesian & Malayo-Polynesian \\
\textcolor{gray}{Filipino Visaya Chabakano} & Creole & Spanish-based \\
\textcolor{teal!75!green}{French} & Indo-European & Italic \\
\textcolor{green!75!teal}{French Basque} & Basque & Basque \\
\textcolor{blue!85!orange}{Great Andamanese} & Andamanese & Great-Andamanese \\
\textcolor{teal!75!green}{Gujarati Houston} & Indo-European & Indo-Iranian \\
\textcolor{lime!75!cyan!90!black}{Hallaki} & Dravidian & Southern-Dravidian \\
\textcolor{yellow!75!lime!90!black}{Han} & Sino-Tibetan & Chinese \\
\textcolor{yellow!75!lime!90!black}{Han BJ} & Sino-Tibetan & Chinese \\
\textcolor{yellow!75!lime!90!black}{Han Cantonese} & Sino-Tibetan & Chinese \\
\textcolor{yellow!75!lime!90!black}{Han Mandarin} & Sino-Tibetan & Chinese \\
\textcolor{yellow!75!lime!90!black}{Han Singapore} & Sino-Tibetan & Chinese \\
\textcolor{purple!75!lime}{Hausa} & Afro-Asiatic & Chadic \\
\textcolor{teal!75!green}{Hazara} & Indo-European & Indo-Iranian \\
\textcolor{olive!66!teal}{Hezhen} & Altaic & Tungusic \\
\textcolor{teal!75!green}{Hindi} & Indo-European & Indo-Iranian \\
\textcolor{orange!66!blue}{Hmong} & Hmong-Mien & Hmongic \\
\textcolor{orange!66!blue}{Hmong Miao} & Hmong-Mien & Hmongic \\
\textcolor{orange!75!blue}{Htin Mal} & Austro-Asiatic & Mon-Khmer \\
\textcolor{red}{Igbo} & Niger-Congo & Atlantic-Congo \\
\textcolor{lime!75!cyan!90!black}{Indian Singapore} & Dravidian & Southern-Dravidian \\
\textcolor{blue!66!orange}{Iraya} & Austronesian & Malayo-Polynesian \\
\textcolor{olive}{Japanese} & Japonic & Japanese \\
\textcolor{olive}{Japanese Tokyo} & Japonic & Japanese \\
\textcolor{blue!66!orange}{Javanese} & Austronesian & Malayo-Polynesian \\
\textcolor{orange}{Jiamao} & Tai-Kadai & Hlai \\
\textcolor{yellow!75!lime!90!black}{Jinuo} & Sino-Tibetan & Tibeto-Burman \\
\textcolor{purple}{Kaba} & Nilo-Saharan & Central-Sudanic \\
\textcolor{teal!75!green}{Kalash} & Indo-European & Indo-Iranian \\
\textcolor{blue!66!orange}{Kambera} & Austronesian & Malayo-Polynesian \\
\textcolor{lime!75!cyan!90!black}{Kamsali} & Dravidian & South-Central-Dravidian \\
\textcolor{yellow!75!lime!90!black}{Karen} & Sino-Tibetan & Tibeto-Burman \\
\textcolor{violet}{Karitiana} & Tupi & Arikem \\
\textcolor{teal!75!green}{Kashmiri Pandit} & Indo-European & Indo-Iranian \\
\textcolor{orange!75!blue}{Kharia} & Austro-Asiatic & Munda \\
\textcolor{red}{Kongo} & Niger-Congo & Atlantic-Congo \\
\textcolor{olive}{Koreans} & Korean & Korean \\
\textcolor{lime!75!cyan!90!black}{Kurumba} & Dravidian & Southern-Dravidian \\
\textcolor{yellow!75!lime!90!black}{Lahu} & Sino-Tibetan & Tibeto-Burman \\
\textcolor{blue!66!orange}{Lamaholot} & Austronesian & Malayo-Polynesian \\
\textcolor{orange!75!blue}{Lawa} & Austro-Asiatic & Mon-Khmer \\
\textcolor{blue!66!orange}{Lembata} & Austronesian & Malayo-Polynesian \\
\textcolor{teal!75!green}{Lodi} & Indo-European & Indo-Iranian \\
\textcolor{red}{Luhya Kenya} & Niger-Congo & Atlantic-Congo \\
\textcolor{purple}{Maasai Kenya} & Nilo-Saharan & Eastern-Sudanic \\
\textcolor{purple!75!lime}{Mada} & Afro-Asiatic & Chadic \\
\textcolor{lime!75!cyan!90!black}{Madiga} & Dravidian & South-Central-Dravidian \\
\textcolor{teal!75!green}{Makrani} & Indo-European & Indo-Iranian \\
\textcolor{lime!75!cyan!90!black}{Mala} & Dravidian & South-Central-Dravidian \\
\textcolor{blue!66!orange}{Malay} & Austronesian & Malayo-Polynesian \\
\textcolor{blue!66!orange}{Malay Singapore} & Austronesian & Malayo-Polynesian \\
\textcolor{blue!66!orange}{Mamanwa} & Austronesian & Malayo-Polynesian \\
\textcolor{red}{Mandenka} & Niger-Congo & Mande \\
\textcolor{blue!66!orange}{Manggarai} & Austronesian & Malayo-Polynesian \\
\textcolor{teal!75!green}{Marathi} & Indo-European & Indo-Iranian \\
\textcolor{violet!85!orange}{Maya} & Mayan & Yuacatecan \\
\textcolor{red}{Mbororo Fulani} & Niger-Congo & Atlantic-Congo \\
\textcolor{purple}{Mbuti Pygmies} & Nilo-Saharan & Central-Sudanic \\
\textcolor{teal!75!green}{Meghawal} & Indo-European & Indo-Iranian \\
\textcolor{cyan!85!blue}{Melanesians Naasioi} & South-Bougainville & Nasioi \\
\textcolor{blue!66!orange}{Mentawai} & Austronesian & Malayo-Polynesian \\
\textcolor{teal!75!green}{Mexican LA} & Indo-European & Italic \\
\textcolor{orange!66!blue}{Miaozu} & Hmong-Mien & Hmongic \\
\textcolor{blue!66!orange}{Minanubu Manobo} & Austronesian & Malayo-Polynesian \\
\textcolor{orange!75!blue}{Mlabri} & Austro-Asiatic & Mon-Khmer \\
\textcolor{orange!75!blue}{Mon} & Austro-Asiatic & Mon-Khmer \\
\textcolor{olive!66!teal}{Mongola} & Altaic & Mongolic \\
\textcolor{purple!75!lime}{Mozabite} & Afro-Asiatic & Berber \\
\textcolor{cyan!85!blue}{NAN Melanesian} & South-Bougainville & Nasioi \\
\textcolor{lime!75!cyan!90!black}{Naidu} & Dravidian & South-Central-Dravidian \\
\textcolor{yellow!75!lime!90!black}{Naxi} & Sino-Tibetan & Tibeto-Burman \\
\textcolor{orange!75!blue}{Negrito Jehai} & Austro-Asiatic & Mon-Khmer \\
\textcolor{orange!75!blue}{Negrito Kensiu} & Austro-Asiatic & Mon-Khmer \\
\textcolor{teal!75!green}{North Italian} & Indo-European & Italic \\
\textcolor{yellow!75!lime!90!black}{Nysha} & Sino-Tibetan & Tibeto-Burman \\
\textcolor{olive}{Okinawan} & Japonic & Ryukyuan \\
\textcolor{blue!85!orange}{Onge} & Andamanese & South-Andamanese \\
\textcolor{teal!75!green}{Orcadian} & Indo-European & Germanic \\
\textcolor{olive!66!teal}{Oroqen} & Altaic & Tungusic \\
\textcolor{teal!75!green}{Pahari} & Indo-European & Indo-Iranian \\
\textcolor{purple!75!lime}{Palestinian} & Afro-Asiatic & Semitic \\
\textcolor{orange!75!blue}{Palaung} & Austro-Asiatic & Mon-Khmer \\
\textcolor{cyan}{Papuan} & Sepik & Ndu \\
\textcolor{teal!75!green}{Pathan} & Indo-European & Indo-Iranian \\
\textcolor{violet!85!orange}{Pima} & Uto-Aztecan & Southern-Uto-Aztecan \\
\textcolor{orange!75!blue}{Plang Blang} & Austro-Asiatic & Mon-Khmer \\
\textcolor{teal!75!green}{Russian} & Indo-European & Slavic \\
\textcolor{teal!75!green}{Sahariya} & Indo-European & Indo-Iranian \\
\textcolor{magenta}{San} & Khoisan & Southern-africa \\
\textcolor{orange!75!blue}{Santhal} & Austro-Asiatic & Munda \\
\textcolor{teal!75!green}{Sardinian} & Indo-European & Italic \\
\textcolor{teal!75!green}{Satnami} & Indo-European & Indo-Iranian \\
\textcolor{orange!66!blue}{She} & Hmong-Mien & Ho-Nte \\
\textcolor{lime!75!cyan!90!black}{Siddi} & Dravidian & Southern-Dravidian \\
\textcolor{teal!75!green}{Sindhi} & Indo-European & Indo-Iranian \\
\textcolor{yellow!75!lime!90!black}{Spiti} & Sino-Tibetan & Tibeto-Burman \\
\textcolor{teal!75!green}{Srivastava} & Indo-European & Indo-Iranian \\
\textcolor{blue!66!orange}{Sunda} & Austronesian & Malayo-Polynesian \\
\textcolor{violet}{Surui} & Tupi & Monde \\
\textcolor{orange}{Tai Khuen} & Tai-Kadai & Kam-Tai \\
\textcolor{orange}{Tai Lue} & Tai-Kadai & Kam-Tai \\
\textcolor{orange}{Tai Yong} & Tai-Kadai & Kam-Tai \\
\textcolor{orange}{Tai Yuan} & Tai-Kadai & Kam-Tai \\
\textcolor{lime!75!cyan!90!black}{Telugu Kannada} & Dravidian & Southern-Dravidian \\
\textcolor{blue!66!orange}{Temuan} & Austronesian & Malayo-Polynesian \\
\textcolor{teal!75!green}{Tharu} & Indo-European & Indo-Iranian \\
\textcolor{blue!66!orange}{Toraja} & Austronesian & Malayo-Polynesian \\
\textcolor{teal!75!green}{Toscani Italia} & Indo-European & Italic \\
\textcolor{olive!66!teal}{Tu} & Altaic & Mongolic \\
\textcolor{yellow!75!lime!90!black}{Tujia} & Sino-Tibetan & Tibeto-Burman \\
\textcolor{teal!75!green}{Tuscan} & Indo-European & Italic \\
\textcolor{olive!66!teal}{Uyghur} & Altaic & Turkic \\
\textcolor{olive!66!teal}{Uygur} & Altaic & Turkic \\
\textcolor{teal!75!green}{Vaish} & Indo-European & Indo-Iranian \\
\textcolor{lime!75!cyan!90!black}{Velama} & Dravidian & South-Central-Dravidian \\
\textcolor{lime!75!cyan!90!black}{Vysya} & Dravidian & South-Central-Dravidian \\
\textcolor{orange!75!blue}{Wa} & Austro-Asiatic & Mon-Khmer \\
\textcolor{red}{Xhosa} & Niger-Congo & Atlantic-Congo \\
\textcolor{olive!66!teal}{Xibo} & Altaic & Tungusic \\
\textcolor{olive!66!teal}{Yakut} & Altaic & Turkic \\
\textcolor{orange!66!blue}{Yao Iu Mien} & Hmong-Mien & Mienic \\
\textcolor{yellow!75!lime!90!black}{Yizu} & Sino-Tibetan & Tibeto-Burman \\
\textcolor{red}{Yoruba} & Niger-Congo & Atlantic-Congo \\
\textcolor{red}{Yoruba Nigeria} & Niger-Congo & Atlantic-Congo \\
\textcolor{orange}{Zhuang N} & Tai-Kadai & Kam-Tai \\
\hline
\end{longtable}}

The colours of the population names in the above table are those that where
used in the graphics. These colours where chosen according to linguistic
affiliations and geography. They were used to distinguish the clusters in
the bar plots (see below).

\subsection{Data analysis and visualization}

The combined dataset was analysed using the program \texttt{frappe}
\citep[][\url{http://med.stanford.edu/tanglab/software/frappe.html}]{Tang_et_al_2005},
with $K$ (number of clusters to use) ranging from $2$ to $16$. The graphics
were produced using a combination of \texttt{python} scripts
and the TikZ/PGF graphic system (\url{http://sourceforge.net/projects/pgf/}).

In the bar plots, each cluster was given the colour of the population which had
the highest proportion of this cluster, except when this rule would have given
the same colour to several clusters. In such cases, the clusters where
differentiated by darker or lighter shades of the common colour. The goal of
these rules was to enable an automatic colour attribution to the clusters. This
was necessary given the large amount of graphics produced. Often (but not
always: see p.~\pageref{colours}), the resulting colour attribution allows the
visual recognition of a cluster across the different values of $K$.

Profile trees\refstepcounter{labs}\label{profile_trees} used for the discussion were built, for a given value of $K$ and
a given selection of populations, by computing the pairwise $\chi^2$ distances
between the vectors representing the average profiles of the populations.
The distance matrix was then used to build a tree with \texttt{fastme}
\citep{Desper_and_Gascuel_2002}. The trees were plotted using a combination of
\texttt{python} scripts and the TikZ/PGF graphic system.

\section*{Conflicts of Interest}

The author declares no conflict of interest.

\section*{Acknowledgements}

Thanks to those who gave me access to the data as well as to the DNA donors.\\
The `HGDP' data are available here:
\url{ftp://ftp.cephb.fr/hgdp_supp1/}\\
The `HapMap' data are available here:
\url{http://www.sanger.ac.uk/humgen/hapmap3/}\\
The `Asia' data were obtained from the PASNP consortium:
\url{http://www4a.biotec.or.th/PASNP/}\\
The `India and `Africa' data were obtained from the authors of \cite{Reich_et_al_2009} and \cite{Bryc_et_al_2010} respectively.\\
The information about language families was retrieved from \url{http://www.ethnologue.com/web.asp}.\\
Thanks to Riccardo Zecchina for giving me the opportunity to work on human
population genetics. Thanks to Rapha\"elle Chaix, Cornelia Di Gaetano, \'Evelyne
Heyer, Floriana Voglino and Jean-Fran\c{c}ois Flot for useful discussion and
encouragement. Thanks to Matthieu Guillaumin for suggesting the use of a
$\chi^2$ distance for the comparison of profiles.\\
Thanks to Mark Hahnel for the FigShare repository.\\
Thanks to the anonymous reviewers who accepted to read and comment an earlier version of this paper when it was submitted to \emph{MDPI Genes}.\\
Thanks to Cymon Cox for a few native speaker advice.\\
The author is greatly indebted to Till Tantau, the author of TikZ and PGF.\\
The author was supported by a grant from the University of Piemonte (Italy).






\bibliographystyle{elsarticle-harv}



\bibliography{journ,biblio}

%% file: Supplement.tex
\section*{Annex: detailed description of the results\refstepcounter{labs}\label{annex}\pdfbookmark{Annex}{annex}}

\subsection*{K = 2\refstepcounter{labs}\label{K2}\pdfbookmark[1]{K=2}{annex:K2}}

Raw results:
\texttt{Frappe\_K2.txt}\footnote{%
\url{http://dx.doi.org/10.6084/m9.figshare.104}}

Profiles of the individuals:
\texttt{Frappe\_K2.pdf}\footnote{%
\url{http://dx.doi.org/10.6084/m9.figshare.276}}

Average profiles of the populations: 
\texttt{Frappe\_K2\_pops.pdf}\footnote{%
\url{http://dx.doi.org/10.6084/m9.figshare.188}}

Ranked average profiles of the populations: 
\texttt{Frappe\_K2\_rankings.pdf}\footnote{%
\url{http://dx.doi.org/10.6084/m9.figshare.291}}

The separation in 2 clusters differentiates between a `Sub-Saharan' trend
(\textcolor{red}{cluster $1$}) and an `East Asian' trend
(\textcolor{blue!66!orange}{cluster $2$}).

The most typically `Sub-Saharan' population is a Bantu population, and the most
typically `East Asian' is an Austronesian population from Taiwan. The Bantu
populations are known for having spread over a large part of Sub-Saharan Africa during
the last millenia and the Austronesians have done the same in the Pacific and
Indian oceans, with a probable origin in Taiwan.

African populations have a large predominance of cluster $1$. The Sub-Saharan
populations with a noticeable component $2$ are the Fulani and the Maasai. The
Fulani are West-African nomads whose origins are controversial. It is sometimes
proposed that they have migrated from more eastern regions of Africa. The
Maasai are an East African population which probably originates from North-East
Africa. Unfortunately, the dataset lacks some populations from Sudan or from
the Horn of Africa.

The proportion of cluster $1$ is partly correlated to distance from Sub-Saharan
Africa, with the following gradient:\\
Sub-Saharan Africa $>$ North Africa $>$ Middle East $>$ Europe $>$ Pakistan $>$ India.

As expected from their African ancestry, Siddi (`African Indians') and African
Americans have high cluster $1$ proportions.

Cluster $1$ is noticeable in populations from America and Oceania. It should be noted
that the Oceanians in the dataset are not Austronesians. It could be
interesting to add some Polynesian populations to the dataset.\\
Non-Taiwanese Austronesians in the dataset are not among those presenting the
highest proportions of cluster $2$. This difference with Taiwanese could be
explained by some admixture between Malayo-Polynesians and other populations
such as Indians in the maritime territories of South-East Asia. In coherence
with this hypothesis is the fact that most continental East and South-East
Asian populations (Sino-Tibetans, Tai-Kadai, Hmong-Mien and some
Austro-Asiatic) show a very high cluster $2$ proportion, like the Taiwanese
Austronesians. The exceptions are Mon and Cambodians, two Austro-Asiatic
populations of Indochina that have a little more cluster $1$ proportion than
the others (but their profile is still predominantly composed by cluster $2$,
and the influence of India has been strong on Indochina too\refstepcounter{labs}\label{suppl:Indochina}).


Altaic populations show various proportions of cluster $1$. In this regard,
they differ from Koreans and Japanese, to whom they are sometimes related by
linguists. Koreans and Japanese have profiles more similar to Sino-Tibetan
populations, i.e. a very low cluster $1$ proportion. This low proportion in
East Asian populations contrasts with what is observed in American
populations. If the ancestry of the latter is to be found somewhere in Asia, it
would probably not be from a stem with a profile similar to that of extant
East Asians. It should be noted that the sample of American populations does
not contain Na-Dene or Eskimo-Aleut speakers. Including the data from
\cite{Rasmussen_et_al_2010} could yield interesting results\refstepcounter{labs}\label{suppl:Altaic-America}.

\subsection*{K = 3\refstepcounter{labs}\label{K3}\pdfbookmark[1]{K=3}{annex:K3}}

Raw results:
\texttt{Frappe\_K3.txt}\footnote{%
\url{http://dx.doi.org/10.6084/m9.figshare.105}}

Profiles of the individuals:
\texttt{Frappe\_K3.pdf}\footnote{%
\url{http://dx.doi.org/10.6084/m9.figshare.277}}

Average profiles of the populations:
\texttt{Frappe\_K3\_pops.pdf}\footnote{%
\url{http://dx.doi.org/10.6084/m9.figshare.95713}}

Ranked average profiles of the populations:
\texttt{Frappe\_K3\_rankings.pdf}\footnote{%
\url{http://dx.doi.org/10.6084/m9.figshare.292}}

The 3 trends are `African' (cluster $1$), `European'
(cluster $2$) and `East Asian'
(cluster $3$).

\textcolor{red}{Cluster $1$} is overwhelming in Sub-Saharan African populations, except for the two
previously noted Fulani and Maasai, which show a significant proportion of
cluster $2$. Among Bantu-speaking population, north-eastern Bantu and Luhya from Kenya
show a little more of cluster $2$ than the others (which is not surprising,
considering the geographic proximity of these populations with the
Maasai\refstepcounter{labs}\label{suppl:Kenya}). The same holds for the Nilo-Saharan-speaking Bulala.
Cluster $1$ is dominant in `African Indians' (Siddi) and African Americans.

Cluster $1$ is important in Mozabites from North Africa, Bedouins and
Palestinians from Middle East. Some Mozabite and Bedouin individuals have
more than $50\%$ cluster $1$.

In places geographically more distant to Africa, cluster $1$ is found with an
important proportion in some individuals in Makrani and Sindhi, populations
from southern Pakistan. This could be explained by admixture with descendants
from African slaves or soldiers (Sheedis) that are established in these
regions.

Cluster $1$ is also noticeable in Oceanian populations, and to various degrees in
some populations of maritime South-East Asia:
\begin{itemize}
\item Onge and Great Andamanese (from the Andaman islands);
\item Jehai and Kensiu (Negritos from Malaysia);
\item Kambera, Manggarai, Lamaholot, Lembata and Alorese (from the Lesser Sunda Islands);
\item Mamanwa, Agta, Ati and Ayta (Negritos from the Philippines).
\end{itemize}
I will use the abbreviation ANLS to designate this group of populations:
Andaman, Negrito, Lesser Sunda. The presence of cluster $1$ in these
populations could be a genetic trace of the ancient colonization of these
regions by an early wave of migration out of Africa\refstepcounter{labs}\label{suppl:early}. It
would be interesting in this regard to add Australian populations to the data,
as Australia is thought to have been reached early in the history of world
colonization by modern humans.

\textcolor{green!75!teal}{Cluster $2$} is predominant in populations from North Africa, Middle East, Europe,
Pakistan and the Dravidian and Indo-European populations of India. There are
however some Indo-European-speaking populations with a somewhat lower cluster $2$
proportion. For example, Hazara from northern Pakistan, who have some Altaic
origins\refstepcounter{labs}\label{suppl:Hazara}, and Himalayan populations (Pahari), who live in close
contact with Sino-Tibetan populations\refstepcounter{labs}\label{suppl:Himalaya}.

Among populations with a high cluster $2$ proportion, those from West and South
Europe have the highest proportion. The cluster $2$ proportion is slightly
lower for populations of the Middle East (who have instead a higher cluster $1$
proportion) and for populations in East-Europe and Pakistan (who have a higher
cluster $3$ proportion). For the populations of India the decrease in cluster
$2$ (`compensated' by an increase in cluster $3$) continues, with a tendency
for Dravidian populations to have a lower cluster $2$ proportion than
Indo-European populations. 

Cluster $2$ is important in American populations and in some Altaic populations
such as Uyghur and Yakut. As for $K=2$, American populations are more similar
in clustering profile to Altaic populations than to other Asian populations. As
noted previously (p.~\pageref{suppl:Altaic-America}), the inclusion of the data from
\cite{Rasmussen_et_al_2010} could be highly interesting, because this study not
only had Na-Dene and Eskimo-Aleut samples, but also a fair variety of Siberian
populations.\\
Cluster $2$ is also important in the Himalayan Sino-Tibetan populations (Spiti).
This observation is coherent with the results from the study of Y chromosomes:
Himalayan Sino-Tibetan populations have a high diversity of Y haplotypes,
indicating complex ancestry \citep{Su_et_al_2000}. The high proportion of cluster
$2$ could for example be explained by an Altaic contribution in Spiti's ancestry.
Some admixture with Indo-Europeans is also probable, given the localisation of
the sampled population (Jammu and Kashmir).

Similarly to cluster $1$, cluster $2$ is noticeable in various populations of
maritime South-East Asia. It is also noticeable in some populations speaking
Austro-Asiatic languages: Kharia and Santhal from India, Cambodians, Mon from
Thailand, Kensiu and Jehai from peninsular Malaysia. Admixture with
neighbouring Indian populations is highly probable in the case of Kharia and
Santal, and the hypothesis of an Indian influence in maritime South-East Asia
proposed for $K=2$ (p.~\pageref{suppl:Indochina}) can be invoked again to explain the
presence of cluster $2$ in the populations of South-East Asia.

\textcolor{orange!66!blue}{Cluster $3$} is highly predominant in Hmong-Mien and Tai-Kadai populations, most
Sino-Tibetan populations, Koreans, Japanese, and some Austronesian populations:
Atayal and Ami (from Taiwan), Bidayuh and Dayak from Borneo, Mentawai (west of
Sumatra), Toraja (from Sulawesi), Manobo and Filipinos (from the Philippines).
More generally, it is by far the main component in all populations from East
and South-East Asia, and constitutes an important part of the clustering profiles
of populations from Oceania, America and Central and North Asia. It decreases
in favor of cluster $2$ following an east $>$ west gradient in populations of
India, Pakistan and East Europe.

\subsection*{K = 4\refstepcounter{labs}\label{K4}\pdfbookmark[1]{K=4}{annex:K4}}

Raw results:
\texttt{Frappe\_K4.txt}\footnote{%
\url{http://dx.doi.org/10.6084/m9.figshare.106}}

Profiles of the individuals:
\texttt{Frappe\_K4.pdf}\footnote{%
\url{http://dx.doi.org/10.6084/m9.figshare.278}}

Average profiles of the populations:
\texttt{Frappe\_K4\_pops.pdf}\footnote{%
\url{http://dx.doi.org/10.6084/m9.figshare.189}}

Ranked average profiles of the populations:
\texttt{Frappe\_K4\_rankings.pdf}\footnote{%
\url{http://dx.doi.org/10.6084/m9.figshare.293}}

Here, an `American' cluster (number $4$) is added to
the three previous ones: `African' (cluster $1$),
`European' (number $2$) and `East Asian'
(number $3$).

Compared to the case where $K=3$, comments regarding the distribution of cluster
$1_3$ apply also to \textcolor{red}{cluster $1_4$}. For \textcolor{teal!75!green}{cluster $2_4$}, the only notable change with
respect to cluster $2_3$ is that American populations loose most of their cluster
$2$ component (this partially affects Mexicans). The same occurs for \textcolor{blue!66!orange}{cluster $3$}.
Altaic, Sino-Tibetan and Hmong-Mien populations also tend to have less cluster
$3$ proportion, but to a lesser extent, while the opposite tendency is observed
for Austronesian, Tai-Kadai and Austro-Asiatic populations. Although it has a
somewhat different distribution from cluster $3_3$, cluster $3_4$ is still the most
prominent cluster for South-East, East and North Asia.

\textcolor{violet}{Cluster $4$} is the main cluster for American populations, particularly for
South Americans. Differences between American populations may reflect various
degrees of European and African ancestry. In other populations, cluster $4$ is
rather low, but more present in Altaic populations, Japanese, Koreans and the
Sino-Tibetan populations from India (Nysha, Aonaga and Spiti), followed by
Hazara, Russians, Pahari, non-Indian Sino-Tibetans, Burusho and Hmong-Mien. It is
absent or almost absent in African populations. 

Not surprisingly, the profile of Mexicans is approximately composed of half
cluster $2$ (putative European ancestry) and half cluster $4$ (putative American ancestry). The
similarity between the Indo-European Hazara and the Altaic Uyghur (see
p.~\pageref{suppl:Hazara}) is reflected by the fact that Hazara are the
Indo-European population with the highest cluster $4$ proportion (after
Mexicans). The relatively high cluster $4$ ranking of Russians might be explained
by some degree of admixture with Siberian populations, and that of Pahari by
admixture with Sino-Tibetan populations (see p.~\pageref{suppl:Himalaya})\refstepcounter{labs}\label{suppl:Himalaya-2}.

To be noted also is the proportion of cluster $4$ in Burusho from northern Pakistan,
which is similar to that of non-Indian Sino-Tibetan populations, and higher
than for the other populations from Pakistan (except Hazara). This population
speaks a language isolate which is sometimes grouped with Sino-Tibetan and
other languages (including some languages spoken in North America) in a
Dene-Caucasian family\refstepcounter{labs}\label{suppl:Burusho}.

\subsection*{K = 5\refstepcounter{labs}\label{K5}\pdfbookmark[1]{K=5}{annex:K5}}

Raw results:
\texttt{Frappe\_K5.txt}\footnote{%
\url{http://dx.doi.org/10.6084/m9.figshare.107}}

Profiles of the individuals:
\texttt{Frappe\_K5.pdf}\footnote{%
\url{http://dx.doi.org/10.6084/m9.figshare.279}}

Average profiles of the populations:
\texttt{Frappe\_K5\_pops.pdf}\footnote{%
\url{http://dx.doi.org/10.6084/m9.figshare.190}}

Ranked average profiles of the populations:
\texttt{Frappe\_K5\_rankings.pdf}\footnote{%
\url{http://dx.doi.org/10.6084/m9.figshare.294}}

Here, there is one cluster for each continent:
\begin{itemize}
\item \textcolor{red}{cluster $1$}, the `African' cluster (more specifically, `Sub-Saharan');
\item \textcolor{teal!75!green}{cluster $2$}, the `European' cluster;
\item \textcolor{yellow!75!lime!90!black}{cluster $3$}, the `Asian' cluster (more specifically, `East Asian');
\item \textcolor{cyan}{cluster $4$}, the `Oceanian' cluster;
\item \textcolor{violet}{cluster $5$}, the `American' cluster;
\end{itemize}

The distribution of cluster $1_5$ is roughly the same as that of cluster $1_4$:
high in African populations. But some interesting differences can be
noticed:\\
The most conspicuous fact is that cluster $1_5$ is almost absent in Oceanian
populations, whereas cluster $1_4$ represented around $8\%$ of their profile.\\
A strong decrease is observed in the ANLS populations, who had been previously
noticed for the presence of cluster $1_3$ (see p.~\pageref{suppl:early}). The relative
decrease is the strongest for the populations of the Lesser Sunda Islands
(Alorese, Kambera, Lamaholot, Lembata, Manggarai), who live the closest to
Oceania and for Kensiu (one of the two Malaysian Negrito populations). The
decrease is also important for the other Negrito populations (Jehai from
Malaysia and Agta, Ati, Ayta and Mamanwa from the Philippines), as well as for
the populations of the Andaman Islands.

Apart from those, most populations outside Africa who had at least a few
percentage points of cluster $1_4$ proportion also have a relatively lower cluster
$1_5$ proportion.\\
The exceptions to this are Sindhi, Makrani, Balochi, Brahui (from Pakistan),
who are affected by a very modest decrease, Siddi and African Americans, who
have a negligible decrease, Mexicans, and populations from the Middle East, for which
the proportion of cluster $1_5$ is even slightly higher than the proportion of
cluster $1_4$.

This observation might suggest means to distinguish between the genetic signature of
recent African ancestry and that pertaining to an ancient out-of-Africa
migration\refstepcounter{labs}\label{suppl:out-of-Africa}. Among populations who had a noticeable cluster
$1$ for $K=3$ and $K=4$, those for which there is no or very little decrease
when considering cluster $1_5$ probably have recent African ancestry. This is
historically known for Siddi and African Americans and probable for Mexicans
also. This was hypothesised for Makrani and Sindhi because of the presence of
descendants from African slaves or soldiers in the south of Pakistan, and it
can be suspected that the same is true for other populations from Pakistan and
Middle East. On the contrary, the populations of Oceania and the ANLS mentioned
p.~\pageref{suppl:early} do not have known recent African ancestry.

Cluster $2_5$ has a distribution very similar to cluster $2_4$. But as in the case of
cluster $1$, cluster $2$ almost completely disappears from the profile of
Oceanians.\\
It also almost disappears from the profiles of the Mlabri
(Austro-Asiatic hunter-gatherers from northern Thailand) and Manggarai, Lembata,
Lamaholot, Kambera and Alorese (Austronesians from the Lesser Sunda
Islands).\\
More generally, there is a relative decrease of cluster $2$ for Austro-Asiatic
and Austronesian populations, as well as for the populations of the Andaman Islands.
The decrease also occurs in Jinuo, Karen, and Tai-Kadai populations but is less
conspicuous because their cluster $2_4$ proportion is already quite low.

At first approximation, cluster $3_4$ seems to have been split between cluster
$3_5$ and cluster $4_5$.\\
Cluster $3_5$ is most important in East Asia. Among the populations with a high
proportions of cluster $3_5$, the rankings according to the importance of this
cluster show a tendency for the following gradient:\\
Chinese and Hmong-Mien $>$ Koreans, Japanese, Taiwanese Austronesians and
Tai-Kadai $>$ Tibeto-Burmese, Mon-Khmer, non-Taiwanese Austronesians and Altaic
populations.

Among non-Taiwanese Austronesians, the lowest proportions of cluster $3$ are
observed in the populations of the Lesser Sunda Islands and the Negritos from
the Philippines (Ayta, Mamanwa, Agta and Ati).

Among the Mon-Khmer-speaking populations, it is lower for the Malaysian
Negritos. It is even lower for the other Austro-Asiatic\footnote{Following the
classification adopted in \cite{Lewis_et_al_2009}, I divide the Austro-Asiatic
populations in two branches: Mon-Khmer (in South-East Asia), and Munda (in
India).} populations, the Kharia and Santhal from India.\\
Cluster $3$ is also an important component of the profile of the
Andamanese populations (Onge and Great Andamanese).

Among Indo-European populations cluster $3$ is important in the profiles of
Pahari, Hazara and Sahariya. I already mentioned (p.~\pageref{suppl:Hazara}) the
Altaic ancestry of the Hazara and the proximity between Pahari and Sino-Tibetan
populations when discussing their low proportion of cluster $2_3$.\\
Apart from Hazara, Burusho (who speak a language isolate) show a higher cluster
$3$ proportion than other populations of Pakistan (see also p.~\pageref{suppl:Burusho}).\\
Among Dravidian populations, some Indians from Singapore show an important
cluster $3$ component. This is probably due to some admixture with Chinese or
Malays.

Papuans have almost exclusively cluster $4_5$, which also constitutes
more than $85\%$ of the profile of Melanesians.\\
It is an interesting fact that the three first non-Oceanian populations in the
ranking according to cluster $4_5$ are Alorese, Lembata and Lamaholot, which are
also those who are geographically the closest to Papua New Guinea.
Apart from populations of the Lesser Sunda Islands, most non-Oceanian
populations with a high proportion of cluster $4_5$ are either Negritos from
Malaysia or the Philippines, Andamanese, or tribal or lower caste
populations from India. These populations from India may bear traces
of an ancient genetic background, pre-dating the arrival of Dravidian and
Indo-European populations.

More generally, cluster $4_5$ is an important component for many populations of
South and South-East Asia, but it tends to be lower for Sino-Tibetan,
Hmong-Mien and Tai-Kadai populations. This distribution is to be related to
the gradient observed for cluster $3_5$. If we set aside Korean, Japanese and
Altaic populations (who have a very low cluster $4_5$ proportion) and populations
from India and Pakistan (who have a low cluster $3_5$ proportion), the
distributions of clusters $3_5$ and $4_5$ are complementary.

Cluster $5_5$ has a distribution similar to cluster $4_4$, but with a slight
increase for most populations of mainland India (the exceptions being Pahari
and the Sino-Tibetan Aonaga, Nysha and Spiti), and with a decrease in
populations of East and South-East Asia. The populations with the highest
proportion of cluster $5_5$ are the same as those for cluster $4_4$: Americans,
followed by Altaic populations.

\subsection*{K = 6\refstepcounter{labs}\label{K6}\pdfbookmark[1]{K=6}{annex:K6}}

Raw results:
\texttt{Frappe\_K6.txt}\footnote{%
\url{http://dx.doi.org/10.6084/m9.figshare.108}}

Profiles of the individuals:
\texttt{Frappe\_K6.pdf}\footnote{%
\url{http://dx.doi.org/10.6084/m9.figshare.280}}

Average profiles of the populations:
\texttt{Frappe\_K6\_pops.pdf}\footnote{%
\url{http://dx.doi.org/10.6084/m9.figshare.191}}

Ranked average profiles of the populations:
\texttt{Frappe\_K6\_rankings.pdf}\footnote{%
\url{http://dx.doi.org/10.6084/m9.figshare.295}}

Here, the `East Asian' cluster $3_5$ is split into a `northern' component (cluster
$3_6$) and a `southern' component (cluster $4_6$).

Clusters \textcolor{red}{$1_6$} and \textcolor{teal!75!green}{$2_6$} have the same distributions as clusters $1_5$ (`African')
and $2_5$ (`European').

\textcolor{olive}{Cluster $3_6$} is most important in Japanese and Koreans. The rankings according
to this cluster reveal the following (approximate) gradient:\\
Japanese and Koreans $>$ Altaic and Sino-Tibetans $>$ Hmong-Mien $>$ Tai-Kadai
$>$ Mon-Khmer (except Mlabri, Jehai and Kensiu) and Austronesians $>$
Andamanese, Burusho, Munda (Kharia and Santhal) and Dravidians $>$ Indo-Iranian and
North American populations.\\
Other populations have a rather low cluster $3_6$ proportion.

Mlabri have almost exclusively \textcolor{orange!75!blue}{cluster $4_6$} in their profile.
There is a tendency towards the following $4_6$ importance gradient:\\
Mon-Khmer and Austronesians $>$ Tai-Kadai $>$ Hmong-Mien $>$ Sino-Tibetans $>$
Andamanese and Munda $>$ Melanesians $>$ Altaic, Koreans and Japanese.

Among Austronesian populations, cluster $4_6$ is lower in the Lesser Sunda
Islands and in the Negritos from the Philippines. Among Sino-Tibetan
populations, cluster $4_6$ is more important in Karen, Lahu and Jinuo,
populations sampled near the western Burmese border\footnote{I will use the
abbreviation JKL\refstepcounter{labs}\label{suppl:JKL} for this group of populations: Jinuo, Karen,
Lahu.}, and less important in Nysha, Aonaga and Spiti, populations sampled in
northern India.

\textcolor{cyan}{Cluster $5_6$} has a distribution similar to cluster $4_5$ (`Oceanian'), but a significant
decrease can be noticed in Austronesian, Mon-Khmer, Tai-Kadai, Sino-Tibetan and
Hmong-Mien populations. The diversification of the `East Asian' clusters
seems to happen at the expense of the `Oceanian' cluster.

\textcolor{violet}{Cluster $6_6$} has a distribution similar to cluster $5_5$
(`American'), but with a decrease in Altaic, Japanese, Korean and Sino-Tibetan
populations, likely related to the appearance of the `northern East Asian'
cluster $3_6$.

\subsection*{K = 7\refstepcounter{labs}\label{K7}\pdfbookmark[1]{K=7}{annex:K7}}

Raw results:
\texttt{Frappe\_K7.txt}\footnote{%
\url{http://dx.doi.org/10.6084/m9.figshare.109}}

Profiles of the individuals:
\texttt{Frappe\_K7.pdf}\footnote{%
\url{http://dx.doi.org/10.6084/m9.figshare.281}}

Average profiles of the populations:
\texttt{Frappe\_K7\_pops.pdf}\footnote{%
\url{http://dx.doi.org/10.6084/m9.figshare.192}}

Ranked average profiles of the populations:
\texttt{Frappe\_K7\_rankings.pdf}\footnote{%
\url{http://dx.doi.org/10.6084/m9.figshare.296}}

The new cluster that appears, number $2_7$, having its highest
frequencies in Dravidian populations, and more generally in India and Pakistan,
represents a `South Asian' tendency. This cluster seems to principally replace
parts of the `European' ($2_6$) and `Oceanian' ($5_6$) clusters.

\textcolor{red}{Cluster $1_7$} is mostly unchanged compared to cluster $1_6$.

The new \textcolor{lime!75!cyan!90!black}{cluster $2_7$} is almost absent from Africa, Oceania and America. A
tiny proportion of the `European' cluster $2_6$ that was detectable in Maya and
some African populations has been replaced by cluster $2_7$, but cluster $2_6$ is
mostly preserved as \textcolor{teal!75!green}{cluster $3_7$} in these populations.

The replacement is more visible for populations of Europe and Middle East,
except that it does not seem to affect Sardinians, and only very lightly Basques.
Populations of Middle East and East Europe are more affected, particularly the
Caucasian Adygei.

For the populations of Pakistan, the proportion of the `Oceanian' cluster
($5_6$, then $6_7$) is greatly reduced. It is replaced by cluster $2_7$,
which also replaces part of cluster $2_6$, so that $2_7$ (`South Asian') and
$3_7$ (`European') are roughly in equal parts. The same observation holds for
Altaic populations, but is less conspicuous because clusters $2_6$ and $5_6$ are less
important.

The same is observed also in India, but resulting in a higher $2_7 / 3_7$
ratio. The proportion of remaining cluster $3_7$ is higher in upper-caste
Indo-Iranian populations and lower in Andamanese, Munda and Tibeto-Burmese
populations.

In East and South-East Asia, $2_6$ is mostly replaced by $2_7$. The `Oceanian'
component ($5_6$, then $6_7$) is also generally affected by the replacement,
but less than in South Asia. Cluster $2_7$ highlights the heterogeneity within
the Malay and Indian populations from Singapore, probably reflecting the
various degrees of Indian ancestry found in the individuals composing these two
populations.

The differences in replacement of the `European' cluster $2_6$ by the
`South Asian' cluster $2_6$ has the following notable effects on the rankings
according to the `European' cluster (now $3_7$):
\begin{itemize}
\item an increase of the ranking of Altaic populations (especially Uyghur),
Hazara, Fulani and Nilo-Saharan populations (especially Maasai);
\item a decrease for Onge, Malaysian Negritos and Munda.
\end{itemize}

\textcolor{olive}{Cluster $4_7$} has the same distribution as the `northern East Asian' cluster $3_6$, but
with a noticeable increase in proportion and rank for Oceanian populations,
Mlabri and Alorese.

\textcolor{orange!75!blue}{Cluster $5_7$} has a distribution similar to the `southern East Asian' cluster $4_6$,
but with an increase in the rankings for most populations of India and a
decrease for Middle East, Europe, Oceania and Japan, and for some
Altaic and Nilo-Saharan speakers.

Following the differential replacement of cluster $5_6$ by the new `South Asian'
cluster $2_7$, the top of the ranking according to the importance of the `Oceanian'
cluster ($5_6$ then $6_7$) becomes clearer:\\
Papuans have their profile almost exclusively contituted by \textcolor{cyan}{cluster $6_7$}, closely followed by
Melanesians. Then, populations from the Lesser Sunda Islands have an important
cluster $6_7$ proportion, which decreases with geographic distance from Papua New
Guinea. The decrease continues with Negritos from the Philippines and
Andamanese, and then other non-Filipino populations from the Philippines, as
well as Toraja from Sulawesi.

\textcolor{violet}{Cluster $7_7$} has the same distribution as cluster $6_6$.

\subsection*{K = 8\refstepcounter{labs}\label{K8}\pdfbookmark[1]{K=8}{annex:K8}}

Raw results:
\texttt{Frappe\_K8.txt}\footnote{%
\url{http://dx.doi.org/10.6084/m9.figshare.110}}

Profiles of the individuals:
\texttt{Frappe\_K8.pdf}\footnote{%
\url{http://dx.doi.org/10.6084/m9.figshare.282}}

Average profiles of the populations:
\texttt{Frappe\_K8\_pops.pdf}\footnote{%
\url{http://dx.doi.org/10.6084/m9.figshare.193}}

Ranked average profiles of the populations:
\texttt{Frappe\_K8\_rankings.pdf}\footnote{%
\url{http://dx.doi.org/10.6084/m9.figshare.297}}


Here, a `non-Niger-Congo' cluster ($2_8$) replaces parts of the previous `African'
($1_7$) and `European' ($3_7$) clusters.

Overall, \textcolor{red}{cluster $1_8$} has a distribution similar to cluster $1_7$. But besides a
general decrease in African populations, a contrast can be observed in the
variation of rankings in European populations: Sardinians undergo a strong
decrease in rankings whereas the rankings of more northern populations
(Orcadians, Russians, and to a lesser extant, north Americans of European
origins and French) increase.

The new \textcolor{purple}{cluster $2_8$} constitutes about one third of the profile of the Maasai
(who speak a Nilo-Saharan language). It is also present in a significant
amount in another Nilo-Saharan-speaking population, the Bulala (but less in the
Kaba), and among speakers of Afro-Asiatic languages, particularly in North
Africa and Middle East. The Kaba (Nilo-Saharan) and the Hausa (Afro-Asiatic)
have little cluster $2_8$, like most Niger-Congo-speaking populations

The Niger-Congo-speaking populations with the highest proportion of cluster $2_8$
are Bantu from the north-east and Luhya from Kenya (two populations who live in
the same region as the Maasai), and the Fulani. This observation may be related to
what had been noticed p.~\pageref{suppl:Kenya} when discussing the presence of the
`European' cluster $2_3$ in African populations.

Outside Africa and Middle East, cluster $2_8$ is above $7\%$ in Italy (including
Sardinia), in the Caucasus (Adygei) and in western Pakistan (Makrani, Brahui
and Balochi). It would be interesting to include data for more populations of
East and North Africa, East Europe and West Asia to get a better view of the
geographic distribution of this cluster.

The `European' \textcolor{green!75!teal}{cluster $5_8$} has roughly the same distribution as cluster $3_7$, but
is partly replaced by cluster $2_8$ in some African populations: Fulani, Maasai,
Luhya and Bantu from the north-east, Mada, Kaba and Bulala (where it completely
disappears).\\
This replacement also affects populations from North Africa, Middle East and
Italy (including Sardinia), Adygei from the Caucasus, Brahui, Makrani and
Balochi from western Pakistan.

The other clusters are mostly unchanged with respect to the case where
$K=7$, with the following correspondences:\\
\begin{tabular}{|c|c|c|c|c|c|}
\hline
Cluster                & \textcolor{lime!75!cyan!90!black}{$3_8$} & \textcolor{olive}{$4_8$} & \textcolor{orange!75!blue}{$6_8$} & \textcolor{cyan}{$7_8$} & \textcolor{violet}{$8_8$} \\
\hline
corresponds to cluster & $2_7$ & $4_7$ & $5_7$ & $6_7$ & $7_7$ \\
\hline
\end{tabular}

\subsection*{K = 9\refstepcounter{labs}\label{K9}\pdfbookmark[1]{K=9}{annex:K9}}

Raw results:
\texttt{Frappe\_K9.txt}\footnote{%
\url{http://dx.doi.org/10.6084/m9.figshare.111}}

Profiles of the individuals:
\texttt{Frappe\_K9.pdf}\footnote{%
\url{http://dx.doi.org/10.6084/m9.figshare.283}}

Average profiles of the populations:
\texttt{Frappe\_K9\_pops.pdf}\footnote{%
\url{http://dx.doi.org/10.6084/m9.figshare.194}}

Ranked average profiles of the populations:
\texttt{Frappe\_K9\_rankings.pdf}\footnote{%
\url{http://dx.doi.org/10.6084/m9.figshare.298}}

Here, the `southern East Asian' cluster which was dominant in Mlabri
($6_8$) is decomposed in two clusters ($6_9$ and $7_9$). There are now 3
`East Asian' clusters:
\begin{itemize}
\item Cluster $4_9$ is more present in Altaic, Korean and Japanese populations.
\item Cluster $6_9$ is more present in Austronesian populations.
\item Cluster $7_9$ is typical of Malaysian Negritos.
\end{itemize}

\textcolor{olive}{Cluster $4_9$} has a similar distribution as cluster $4_8$, but with the following
changes in the rankings:
\begin{itemize}
\item a decrease for Mlabri, Oceanians, and some Austronesian populations;
\item an increase for Kensiu (a Malaysian Negrito population), Andamanese, the
Himalayan Spiti and Pahari, Srivastata, Hazara, Uyghur, Yakut, Russians, Burusho,
North Americans and Colombians.
\end{itemize}

\textcolor{blue!66!orange}{Cluster $6_9$} replaces parts of clusters $4_8$ (`northern East Asian') and $6_8$
(`southern East Asian'). This replacement most strongly affects Austronesians,
but the Negritos from the Philippines and the populations from the Lesser Sunda
Islands have less of this cluster than other Austronesians.\\
Cluster $6_9$ is important also in Mon-Khmer (particularly in Mlabri and Ht'in
Mal, but not in Malaysian Negritos), Tai-Kadai, Hmong-Mien and Sino-Tibetan
populations. Whithin these populations, Tai-Kadai tend to have a higher cluster
$6_9$ proportion, and Sino-Tibetans tend to have a lower proportion. Cluster $6_9$ is
found in Koreans, Japanese, Altaic, Melanesians, and some populations of India
(most noticeably in Munda).

\textcolor{orange!75!blue}{Cluster $7_9$} constitutes a large majority of the profile of Malaysian
Negritos. It is found at a significant level in various South and South-East
Asian populations, with the populations of the Andaman islands and a majority of
Austro-Asiatic speakers among the first populations in the rankings.

Little change occurs for `African' (\textcolor{red}{$1$} and \textcolor{purple}{$2$}), `South Asian' (\textcolor{lime!75!cyan!90!black}{$3$}),
`Oceanian' ($7_8$ then \textcolor{cyan}{$8_9$}) and `American' ($8_8$ then \textcolor{violet}{$9_9$}) clusters, except
for a significant decrease in the rankings of Malaysian Negritos.

The `European' (\textcolor{green!75!teal}{$5$}) cluster is mostly unchanged, except for a decrease in the
rankings of Munda and some Dravidian populations.

\subsection*{K = 10\refstepcounter{labs}\label{K10}\pdfbookmark[1]{K=10}{annex:K10}}

Raw results:
\texttt{Frappe\_K10.txt}\footnote{%
\url{http://dx.doi.org/10.6084/m9.figshare.112}}

Profiles of the individuals:
\texttt{Frappe\_K10.pdf}\footnote{%
\url{http://dx.doi.org/10.6084/m9.figshare.284}}

Average profiles of the populations:
\texttt{Frappe\_K10\_pops.pdf}\footnote{%
\url{http://dx.doi.org/10.6084/m9.figshare.195}}

Ranked average profiles of the populations:
\texttt{Frappe\_K10\_rankings.pdf}\footnote{%
\url{http://dx.doi.org/10.6084/m9.figshare.299}}

Mlabri have now their profile exclusively composed of cluster $7_{10}$.
This could be due to the low genetic diversity of this population. Indeed,
Mlabri seem to have undergone a fairly recent founding effect \citep{Oota_et_al_2005}.

\textcolor{orange!75!blue}{Cluster $7_{10}$} partly substitutes the `Austronesian' and `southern East Asian'
clusters $6_9$ (then $6_{10}$) and $7_9$ (then $8_{10}$). This substitution can
be evidenced by considering the populations for which the decreases in the
`Austronesian' and `southern East Asian' clusters are the highest.\\
Decrease in the `Austronesian' cluster:
\begin{itemize}
\item more than 8 points for Mlabri, Ht'in Mal;
\item more than 7 points for Temuans;
\item more than 6 points for Plang Blang, Wa;
\item more than 5 points for Jinuo, Karen, Cambodians, Lawa, Palaung;
\item more than 4 points for Bidayuh, Dayak, Javanese, Sunda, Tai Yuan;
\item more than 3 points for Aonaga, Nysha, Lahu, Santhal, Mon, Malays from Singapore, Dai, Tai Khuen, Tai Yong, Tai Lue, Zhuang;
\item more than 2 points for Satnami, Kharia, Hmong, Iu Mien, Ayta, Malays, Hakka, Tujia, Jiamao.
\end{itemize}
Decrease in the `southern East Asian' cluster:
\begin{itemize}
\item more than 5 points for Malbri;
\item more than 4 points for Ht'in Mal;
\item more than 3 points for Temuans, Plang Blang, Wa;
\item more than 2 points for Pedi, Javanese, Sunda, Jinuo, Karen, Cambodians, Lawa, Palaung.
\end{itemize}
This is correlated with the head of the rankings according to the importance of cluster $7_{10}$.\\
Apart from the Mlabri, whose case has been already discussed, the populations
with the highest proportions of cluster $7_{10}$ are the other non-Negrito
Mon-Khmer populations (Ht'in Mal, Plang Blang, Wa, Lawa, Cambodians, Palaung,
Mon), the Tibeto-Burmese populations sampled near the Burmese border (JKL,
see p.~\pageref{suppl:JKL}), the Tai-Kadai populations, and the Austronesian
populations from the Malaysian peninsula, Java and Borneo.

Except for the decreases mentioned above, the distribution of clusters \textcolor{blue!66!orange}{$6_{10}$} and
\textcolor{orange!75!blue!63}{$8_{10}$} are fairly similar to those of clusters $6_9$ (`Austronesian') and
$7_9$ (`Malaysian Negrito') respectively.

The other clusters are mostly unchanged with respect to the case where
$K=9$, with the following correspondences:\\
\begin{tabular}{|c|c|c|c|c|c|c|c|}
\hline
Cluster                & \textcolor{red}{$1_{10}$} & \textcolor{purple}{$2_{10}$} & \textcolor{lime!75!cyan!90!black}{$3_{10}$} & \textcolor{olive}{$4_{10}$} & \textcolor{green!75!teal}{$5_{10}$} & \textcolor{cyan}{$9_{10}$} & \textcolor{violet}{$10_{10}$}\\
\hline
corresponds to cluster & $1_9$    & $2_9$    & $3_9$    & $4_9$    & $5_9$    & $8_9$    & $9_9$\\
\hline
\end{tabular}

\subsection*{K = 11\refstepcounter{labs}\label{K11}\pdfbookmark[1]{K=11}{annex:K11}}

Raw results:
\texttt{Frappe\_K11.txt}\footnote{%
\url{http://dx.doi.org/10.6084/m9.figshare.113}}

Profiles of the individuals:
\texttt{Frappe\_K11.pdf}\footnote{%
\url{http://dx.doi.org/10.6084/m9.figshare.285}}

Average profiles of the populations:
\texttt{Frappe\_K11\_pops.pdf}\footnote{%
\url{http://dx.doi.org/10.6084/m9.figshare.196}}

Ranked average profiles of the populations:
\texttt{Frappe\_K11\_rankings.pdf}\footnote{%
\url{http://dx.doi.org/10.6084/m9.figshare.300}}

The `African' putative ancestry is now divided in 3 clusters. A new
`Khoisan-Pygmy' cluster is added to the previously identified `general
Sub-Saharan' and `East African-West Asian' clusters.

\textcolor{red}{Cluster $1$} (`general Sub-Saharan') undergoes an important decrease in Pygmies and
San (more than $40$ percentage points). A decrease is also observable in other
African populations, most notably in south-eastern Bantu populations (Pedi,
Tswana, Xhosa, Sotho, Zulu).\\
Outside Africa, a decrease in cluster $1$ is noticeable in Negritos from the Philippines.

\textcolor{purple!63}{Cluster $2_{11}$} is present mainly in African populations. It reaches its highest
proportions in Mbuti Pygmies ($72.10\%$), San ($67.58\%$) and Biaka Pygmies
($52.24\%$). The next populations according to the importance of this cluster are
Bantu populations from south-eastern Africa (Pedi, Tswana, Xhosa, Sotho, Zulu). This
is probably a consequence of genetic exchanges between Khoisan and Bantu
populations in this region \citep[see][]{Schuster_et_al_2010}.\\
It should be noticed that, in the rankings according to cluster $2_{11}$, the
first two populations without obvious African origins are Ayta and Agta, two of
the populations mentioned p.~\pageref{suppl:early} about a possible genetic trace of
an early out-of-Africa migration in the populations of maritime South-East
Asia.\\
It may be interesting in this regard to consider the proportion of cluster
$2_{11}$ with respect to the total of the three `African' clusters $1_{11}$,
$2_{11}$ and $3_{11}$:

Populations from the Lesser Sunda Islands:
\begin{itemize}
\item Kambera $76.26\%$
\item Lamaholot $59.89\%$
\item Manggarai $55.46\%$
\item Lembata $50.13\%$
\item Alorese $31.35\%$
\end{itemize}

Negritos from the Philippines:
\begin{itemize}
\item Ayta $78.39\%$
\item Agta $65.64\%$
\item Mamanwa $57.32\%$
\item Ati $49.54\%$
\end{itemize}

Malaysian Negritos:
\begin{itemize}
\item Jehai $77.11\%$
\item Kensiu $23.60\%$
\end{itemize}

Andamanese:
\begin{itemize}
\item Onge $42.31\%$
\item Great Andamanese $11.84\%$
\end{itemize}

Known Sub-Saharan ancestry in historical times (through African slaves or soldiers):
\begin{itemize}
\item Siddi $9.14\%$
\item African Americans $6.41\%$
\end{itemize}

Probable Sub-Saharan ancestry (same reasons as above, at least for some individuals):
\begin{itemize}
\item Sindhi $15.30\%$
\item Makrani $12.63\%$
\end{itemize}

Possible Sub-Saharan ancestry (through African slaves or soldiers, or because
of geographical proximity with the above-mentioned populations):
\begin{itemize}
\item Mexicans $20.46\%$
\item Brahui $10.27\%$
\item Balochi $9.69\%$
\item Palestinians $6.16\%$
\item Druze $6.09\%$
\item Bedouins $3.23\%$
\item Mozabites $4.33\%$
\end{itemize}

Bantu populations from southern Africa (possible Khoisan ancestry):
\begin{itemize}
\item Pedi $26.03\%$
\item Tswana $25.37\%$
\item Xhosa $19.90\%$
\item Sotho $19.19\%$
\item Zulu $15.11\%$
\item Herero $9.88\%$
\item Ovambo $4.02\%$
\end{itemize}

Khoisan and Pygmies:
\begin{itemize}
\item Mbuti Pygmies $72.27\%$
\item San $68.24\%$
\item Biaka Pygmies $52.66\%$
\end{itemize}

The other Sub-Saharan populations have this proportion ranging from $2.59\%$
(Yoruba) to $11.71\%$ (Maasai). This proportion cannot be reasonably evaluated
in Papuans and Melanesians because the cumulated proportion of their profile representing putative African ancestry is too low (one
Melanesian sample is at $99.99\%$ and the other at $0.58\%$, but they are both
supposed to be taken from the same population).

Except for Great Andamanese and Kensiu, the populations previously hypothesized
to bear the trace of an ancient out-of-Africa migration (ANLS) have more than
$30\%$ of their total `African ancestry' represented by cluster $2_{11}$. Among
African populations or populations with known or suspected African ancestry,
only Pygmies and San have this proportion higher than $30\%$. Great Andamanese
and Kensiu still have a higher relative proportion of cluster $2_{11}$
than the Sub-Saharan populations without suspected Khoisan admixture.

This suggests a scenario in which one or more populations from the same stock
as Khoisan and Pygmies migrated to South-East Asia, and that the Negritos from
Malaysia and the Philippines and the populations of the Andaman and Lesser
Sunda Islands are partially descendants of these populations.\refstepcounter{labs}\label{suppl:hunter-gatherers}\\
The observations on the variations in the `African' cluster when the
`Oceanian' cluster first appeared may be related to this (see
p.~\pageref{suppl:out-of-Africa}).

\textcolor{purple}{Cluster $3_{11}$} corresponds to cluster $2_{10}$, but there is a tendency for the
rankings of San, Pygmies, south-eastern Bantu and ANLS populations to decrease.

The other clusters are mostly unchanged with respect to the case where
$K=10$, with the following correspondences:\\
\begin{tabular}{|c|c|c|c|c|c|c|c|c|}
\hline
Cluster                & \textcolor{lime!75!cyan!90!black}{$4_{11}$} & \textcolor{olive}{$5_{11}$} & \textcolor{green!75!teal}{$6_{11}$} & \textcolor{blue!66!orange}{$7_{11}$} & \textcolor{orange!75!blue}{$8_{11}$} & \textcolor{orange!75!blue!63}{$9_{11}$} & \textcolor{cyan}{$10_{11}$} & \textcolor{violet}{$11_{11}$}\\
\hline
corresponds to cluster & $3_{10}$ & $4_{10}$ & $5_{10}$ & $6_{10}$ & $7_{10}$ & $8_{10}$ & $9_{10}$  & $10_{10}$\\
\hline
\end{tabular}

\subsection*{K = 12\refstepcounter{labs}\label{K12}\pdfbookmark[1]{K=12}{annex:K12}}

Raw results:
\texttt{Frappe\_K12.txt}\footnote{%
\url{http://dx.doi.org/10.6084/m9.figshare.114}}

Profiles of the individuals:
\texttt{Frappe\_K12.pdf}\footnote{%
\url{http://dx.doi.org/10.6084/m9.figshare.286}}

Average profiles of the populations:
\texttt{Frappe\_K12\_pops.pdf}\footnote{%
\url{http://dx.doi.org/10.6084/m9.figshare.197}}

Ranked average profiles of the populations:
\texttt{Frappe\_K12\_rankings.pdf}\footnote{%
\url{http://dx.doi.org/10.6084/m9.figshare.301}}

The `Khoisan-Pygmy' cluster disappears. The comparisons shall therefore
be made with the situation at $K=10$.

A rearrangement of the `East Asian' clusters occurs:
\begin{itemize}
\item There are 2 `Austronesian' clusters ($6_{12}$ and $7_{12}$), one
of which ($6_{12}$) is in fact more specific to the non-Filipino populations of
the Philippines. Cluster $7_{12}$ has a reinforced Austronesian character.
\item A `continental South-East Asian' cluster appears.
\item The `northern East Asian' cluster $4$ acquires a more `maritime' aspect.
\item The `Mlabri-specific' and `Malaysian Negrito-specific' clusters are
maintained.
\end{itemize}

The `African' clusters \textcolor{red}{$1$} and \textcolor{purple}{$2$} and the `European' \textcolor{green!75!teal}{cluster $5$} do not change
much, the most notable difference with respect to the case where $K=10$ is a
decrease in the rankings for Mamanwa.

The distribution of the `Indian' \textcolor{lime!75!cyan!90!black}{cluster $3$} is mostly unchanged. A tendency
towards a decrease in the rankings can be observed for the populations of the
Philippines (especially in Mamanwa), Taiwan and Japan.\refstepcounter{labs}\label{suppl:Japan-Austronesian-1}

The `northern East Asian' \textcolor{olive}{cluster $4$} undergoes a significant decreases in many
Asian populations: Sino-Tibetans, Hmong-Mien, Mon-Khmer (except Mlabri and
Malaysian Negritos), Altaic populations, Pahari, Koreans, Tai-Kadai, Hazara,
Japanese, Sahariya. Among these populations, the decrease tends to be lower in
Japanese, Tai-Kadai and southern Chinese populations. Cluster $4$ increases in
some Austronesian populations. These differences lead to an increased
contrast between populations of Japan and the other populations of northern
East Asia. The rankings of Filipinos and Austronesian Taiwanese increase.

\textcolor{blue!66!orange}{Cluster $6_{12}$} represents about two thirds of the profile of Mamanwa,
nomadic Negritos from the Philippines living in the north of Mindanao.
It also represents more than $8\%$ of the profiles of the other
non-Filipino populations of the Philippines (Ati, Ayta, Agta, Iraya, Manobo).

\textcolor{blue!66!orange!63}{Cluster $7_{12}$} corresponds to the `Austronesian' cluster $6_{10}$, but with
significant changes. A decrease is observed for many populations of Central and
East Asia. The decrease in percentage points is more important in Mamanwa,
Hmong-Mien, Mon-Khmer (except Mlabri and Malaysian Negritos), JKL and
Tai-Kadai. This decrease is still significant in populations in which the
proportion of cluster $6_{10}$ was not very high. This results in a strong
relative decrease for the Sino-Tibetan populations of India (Aonaga, Nysha and
Spiti), Pahari, Kashmiri, Hazara, and Altaic populations. An increase can be
noted in Okinawans. These variations reveal a contrast between `continental' and
`maritime' populations.

The Austronesian populations are more grouped in the top of the rankings
according to cluster $7_{12}$ than they were for cluster $6_{10}$: The first $21$
positions are occupied by Austronesian populations, and they are all found in
the $38$ first positions. Tai-Kadai are the second group of populations
according to the importance of cluster $7_{12}$. They rank between $22$ and
$34$.  It should be noted in this regard that it has been proposed
that Tai-Kadai languages are part of the Austronesian
family \citep{Sagart_2004}\refstepcounter{labs}\label{suppl:Austro-Tai}. Cambodians are the non-Austronesian and non-Tai-Kadai
population with the highest proportion of cluster $7_{12}$. This could be
explained by a possible admixture with Cham, an Austronesian population which
once occupied part of southern Indochina, and which is still present in
Cambodia\refstepcounter{labs}\label{suppl:Cham}, or even by the presence of Cham people in the Cambodian sample.

\textcolor{orange!75!blue}{Cluster $8_{12}$} is similar to the `Mlabri-specific' cluster $7_{10}$, but with
an notable relative decrease for Hmong-Mien, Pahari and Tibeto-Burmese from
continental south China (Naxi, Yizu, Lahu) and north-east India (Aonaga,
Nysha).

\textcolor{orange!75!blue!63}{Cluster $9_{12}$} corresponds to the `Malaysian Negrito-specific' cluster $8_{10}$,
but with an important rank decrease for Mamanwa.

\textcolor{orange!66!blue}{Cluster $10_{12}$} constitutes an important proportion of the profiles of
populations of East Asia. The following approximate cluster $10_{12}$ gradient
shows a `southern continental' $>$ `northern maritime' tendency within East Asia:\\
Hmong-Mien (except She), Tibeto-Burmese (except Spiti) and Palaungic (Lawa,
Palaung, Wa, Plang Blang) Mon-Khmer $>$ Ht'in Mal and Tai-Kadai (except Zhuang)
$>$ She, Chinese and Zhuang $>$ Mon, Cambodians, Tungusic (Hezhen, Xibo,
Oroqen) and Mongolic (Tu, Mongola, Daur) Altaic, Pahari, Spiti and Koreans $>$
Austronesian populations of Java, the Malaysian peninsula and Borneo, Turkic
(Yakut and Uyghur) Altaic, Hazara, Sahariya and Japanese.


\textcolor{cyan}{Cluster $11_{12}$} corresponds to the `Oceanian' cluster $9_{10}$, but with a
decrease for Negritos from the Philippines and important rank decreases in some
populations of Sumatra, Taiwan, the Philippines, and Japan.

\textcolor{violet}{Cluster $12_{12}$} corresponds to the `American' cluster
$10_{10}$. A decrease occurs for Ami and Atayal from Taiwan and Mamanwa
and Iraya from the Philippines.

\subsection*{K = 13\refstepcounter{labs}\label{K13}\pdfbookmark[1]{K=13}{annex:K13}}

Raw results:
\texttt{Frappe\_K13.txt}\footnote{%
\url{http://dx.doi.org/10.6084/m9.figshare.115}}

Profiles of the individuals:
\texttt{Frappe\_K13.pdf}\footnote{%
\url{http://dx.doi.org/10.6084/m9.figshare.287}}

Average profiles of the populations:
\texttt{Frappe\_K13\_pops.pdf}\footnote{%
\url{http://dx.doi.org/10.6084/m9.figshare.198}}

Ranked average profiles of the populations:
\texttt{Frappe\_K13\_rankings.pdf}\footnote{%
\url{http://dx.doi.org/10.6084/m9.figshare.302}}

At $K=13$, there are several important changes:
\begin{itemize}
\item The `Khoisan-Pygmy' cluster observed at $K=11$ reappears ($2_{11}$ then $2_{13}$).
\item A new `Middle Eastern' cluster ($4_{13}$) appears.
\item The cluster specific to the Negritos from the Philippines ($6_{12}$) disappears.
\end{itemize}
The results shall thus be compared to the situation at $K=11$.

\textcolor{red}{Cluster $1_{13}$} corresponds to cluster $1_{11}$. It decreases in African
populations, particularly in the Nilo-Saharan-speaking Maasai and Bulala, but
also in Kaba (who also are Nilo-Saharan speakers), and in the two East African
Niger-Congo populations Luhya and Bantu from the north-east (see
p.~\pageref{suppl:Kenya}), as well as in the Afro-Asiatic Mada. A less important
decrease occurs for the Onge from the Andaman Islands, but this leads to a
very strong effect in terms of relative decrease and rankings.

\textcolor{purple!63}{Cluster $2_{13}$} corresponds to cluster $2_{11}$. An important rank decrease can be
noted in Vaish, Onge, Russians and Kamsali, and an increase in Druze.

\textcolor{purple}{Cluster $3_{13}$} roughly corresponds to cluster $3_{11}$ (it is present mainly in
East and North Africa and Middle East) but is now less important in populations
from West Asia, North Africa and Europe.\\
The `Sub-Saharan' character of cluster $3_{13}$ is reinforced with respect to
cluster $3_{11}$ because important decreases occur for many populations,
particularly in Middle East, North Africa, Europe (especially in Sardinia,
southern Italy and in the Caucasus), and Pakistan. Simultaneously, most
Sub-Saharan populations undergo an increase in cluster $2$. Notable exceptions are
Zulu and Ovambo, two Bantu populations from southern Africa, and Fulani, for
which there is a notable decrease.

The new `Middle Eastern' cluster (\textcolor{purple!75!lime}{$4_{13}$}) constitutes about one third of the
profiles of the populations of Middle East. It is also important for
the populations of western Pakistan (Brahui, Makrani and Balochi), the Adygei
(Caucasus), the Mozabites (North Africa), and the Kalash (more than $15\%$ in
these populations). It is also present at a significant level in the other
populations of Pakistan, in Kashmiri and in the populations of Italy (including
Sardinia),

\textcolor{lime!75!cyan!90!black}{Cluster $5_{13}$} corresponds to the `South Asian' cluster $3_{11}$. A slight
increase can be noted in West and North European populations.

\textcolor{olive}{Cluster $6_{13}$} corresponds to the `northern East Asian' cluster $4_{11}$. A
decrease occurs in southern and continental populations. The decrease has the
following approximate importance gradient:\\
Tibeto-Burmese and Palaungic Mon-Khmer
$>$ Altaic (except Uyghur), Pahari, Ht'in Mal, Hmong-Mien
$>$ Mon, Chinese and Koreans
$>$ Tai-Kadai and Cambodians
$>$ populations of Japan, Hazara and Uyghur
$>$ populations of Java.

\textcolor{green!75!teal}{Cluster $7_{13}$} corresponds to the `European' cluster $6_{11}$. A general decrease
is observed, which is more important in populations from the Middle East (more
than $12$ percentage points lost in these populations). The contrast between
non-Caucasian Europeans and other populations is reinforced because the new
cluster $4_{13}$ replaces a more important part of the `European' cluster
in Adygei and populations from Middle East, North Africa and Pakistan than in
non-Caucasian European populations. Non-Caucasian Europeans have more than
$67\%$ cluster $7_{13}$, the Adygei are at $51.7\%$, and the other populations
are below $50\%$. The proportion of the `European' cluster remains above
$20\%$ in Middle East, North Africa and Pakistan, as well as in Kashmiri,
Uyghur and Mexicans.

\textcolor{blue!66!orange}{Cluster $8_{13}$} is similar to the `Austronesian' cluster $7_{11}$, but with a
significant decrease in many populations of East Asia, most notably in
Mon-Khmer (except Malaysian Negritos and Mlabri), Sino-Tibetans (except Spiti),
Austronesian populations of Java, Borneo and the Malaysian peninsula,
Tai-Kadai and Hmong-Mien.
Within these populations the following contrasts can be noted:
\begin{itemize}
\item Among Mon-Khmer populations, the decrease is stronger in Ht'in Mal and
Palaungic.
\item Among Sino-Tibetans, the decrease is stronger in non-Spiti Tibeto-Burmese,
especially in JKL, and less important in northern Chinese.
\item Among Tai-Kadai, the decrease is slightly less strong in the eastern
populations (Jiamao and Zhuang).
\item Among Hmong-Mien, the decrease is less strong in She.
\end{itemize}
A slight increase occurs in Onge and Mamanwa.

The decreases in the `Austronesian' cluster correlate quite well with the
appearance of a `general southern East Asian' cluster (\textcolor{orange!75!blue!75}{$9_{13}$}). This
cluster accounts for almost one third of the profiles of Palaungic,
Ht'in Mal, and JKL populations. It is present at more than $7\%$ in
Austro-Asiatic (except Malbri and the Kensiu Malaysian Negritos), Hmong-Mien,
Sino-Tibetans, Tai-Kadai, Austronesians from Java, Borneo, the Malaysian
peninsula and Sumatra (except Mentawai), Altaic, Koreans, Pahari and Sahariya.
Contrasts similar as above are visible:
\begin{itemize}
\item Cluster $9_{13}$ is more important in Palaungic and Ht'in Mal than in the
other Mon-Khmer populations.
\item Among Sino-Tibetans, it is more important in non-Spiti Tibeto-Burmese
(especially in JKL) than in Chinese, and it is less important in Spiti.
\item Among Tai-Kadai, it is more important in western populations.
\item Among Hmong-Mien, it is less important in She.
\item The importance of cluster $9_{13}$ is quite variable within Austronesian populations.
It is more important in Temuans (from the Malaysian peninsula) and in the populations of Java.
\item Among Altaic populations, it is less important in the Turkic Yakut and Uyghur.
\end{itemize}

\textcolor{orange!75!blue!50}{Cluster $10_{13}$} corresponds to the `Mlabri-specific' cluster $8_{11}$. A decrease
can be observed, which also correlates with the appearance of cluster $9_{13}$.
It is stronger in Ht'in Mal and Palaungic Mon-Khmer, JKL and Temuans (more than
$2.5$ percentage points).\\
A slight increase can be noticed in some populations of Taiwan and the
Philippines, in Japan and in Mentawai.\refstepcounter{labs}\label{suppl:Japan-Austronesian-2}

\textcolor{orange!75!blue}{Cluster $11_{13}$} corresponds to the `Malaysian Negrito' cluster $9_{11}$. In a
similar way as above, a decrease occurs in the populations that have an
important proportion of cluster $9_{13}$, particularly in Ht'in Mal and Palaungic
Mon-Khmer, JKL, Temuans, Bidayuh (from Borneo) and the populations of Java
(more than $4.5$ percentage points).\\
An increase occurs in populations of Japan, the Philippines, Taiwan, Sulawesi
and in Mentawai.

\textcolor{cyan}{Cluster $12_{13}$} corresponds to the `Oceanian' cluster $10_{11}$. A decrease
occurs in many Austronesian populations (particularly in the Philippines,
less in Java), in Melanesians, Onge and Okinawans. The rankings of Taiwanese
Austronesians and Mentawai strongly decreases.\refstepcounter{labs}\label{suppl:Okinawa-Andaman-1}

\textcolor{violet}{Cluster $13_{13}$} corresponds to the `American' cluster $11_{11}$. A decrease
occurs in Ami from Taiwan and in Indo-European (except Pahari and populations
from Pakistan), Dravidian (except Brahui from Pakistan) and Munda populations.

\subsection*{K = 14\refstepcounter{labs}\label{K14}\pdfbookmark[1]{K=14}{annex:K14}}

Raw results:
\texttt{Frappe\_K14.txt}\footnote{%
\url{http://dx.doi.org/10.6084/m9.figshare.116}}

Profiles of the individuals:
\texttt{Frappe\_K14.pdf}\footnote{%
\url{http://dx.doi.org/10.6084/m9.figshare.287}}

Average profiles of the populations:
\texttt{Frappe\_K14\_pops.pdf}\footnote{%
\url{http://dx.doi.org/10.6084/m9.figshare.199}}

Ranked average profiles of the populations:
\texttt{Frappe\_K14\_rankings.pdf}\footnote{%
\url{http://dx.doi.org/10.6084/m9.figshare.303}}

The `Middle Eastern' cluster disappears, but the `Khoisan-Pygmy' cluster is still
there. Therefore, for the `African' clusters, the comparisons will be
made with the situation at $K=11$, which is probably quite similar.

The Asian clusters are highly reorganized:
\begin{itemize}
\item There are two `Austronesian' clusters. Cluster $7_{14}$ is dominant
in Borneo, Java and the Malaysian peninsula and cluster $8_{14}$ is dominant in
the Philippines.
\item There is a `southern East Asian' cluster ($11_{14}$) predominant
in Hmong-Mien and Sino-Tibetan populations.
\item There is a cluster specific to the Andamanese and Negritos from
the Philippines ($12_{14}$).
\item The `Indian' ($4_{14}$), `northern East Asian' ($5_{14}$),
`Mlabri-specific' ($9_{14}$), and `Malaysian Negrito' ($10_{14}$) clusters can
still be identified.
\end{itemize}

\textcolor{red}{Cluster $1_{14}$} corresponds to the `general Sub-Saharan' cluster $1_{11}$. The only
important difference is that it disappears from the profile of Onge.

\textcolor{purple!63}{Cluster $2_{14}$} is similar to the `Khoisan-Pygmy' cluster $2_{11}$. It disappears
from the profile of Onge and decreases in Great Andamanese, in the
Negritos from the Philippines and in some populations of India.

\textcolor{purple}{Cluster $3_{14}$} corresponds to the `East African-West Asian' cluster $3_{11}$.  It
disappears from the profile of Onge, and also slightly decreases in
Great Andamanese, Sardinians, and in the populations of Middle East and North Africa.

\textcolor{lime!75!cyan!90!black}{Cluster $4_{14}$} is similar to the previously described `Indian' cluster.
It constitutes the majority of the profiles of most Dravidian
populations. The exceptions are Brahui from Pakistan ($38.99\%$) and the
`African Indians' Siddi ($16.21\%$). It can be noted that the Indians from
Singapore have a somewhat lower cluster $4_{14}$ proportion compared to the
Dravidian populations of India. This could be explained by some admixture with
Chinese or Malay populations.\\
Cluster $4_{14}$ is also important in other populations of India and Pakistan.
It is above $50\%$ in the Indo-Iranian populations of India except Sahariya
($48.47\%$), Kashmiri ($45.41\%$) and Pahari ($27.09\%$). It is important in
Munda and still notable in Great Andamanese and Spiti. In Pakistan the
proportion of cluster $4_{14}$ is highest in Sindhi ($44.68\%$) and lowest in
Hazara ($17.39\%$). Outside Pakistan and India, cluster $4_{14}$ is notable in
Adygei, Uyghur and Mon. This presence in Mon could be related to the long time
period when Indochina received commercial, political and cultural inputs from
India and Sri Lanka (see also p.~\pageref{suppl:Indochina}).

\textcolor{olive}{Cluster $5_{14}$} is similar to the previously described `northern East Asian'
cluster. However, it displays a clear contrast between the populations of
Japan and the other populations. This seems stronger than the contrast already
observed at $K=12$.
Cluster $5_{14}$ constitutes almost $75\%$ of the profile of Okinawans,
almost $65\%$ in Japanese and almost $50\%$ in Koreans. It then decreases
according to the following approximate gradient:\\
Altaic (except Uyghur)
$>$ Sino-Tibetans (except Spiti, southern Chinese and JKL)
$>$ southern Chinese, Spiti, She, Hazara, Uyghur and Pahari
$>$ JKL, Miaozu, Iu Mien, Palaungic, Mon, Cambodians, Filipinos and Austronesian Taiwanese.

\textcolor{green!75!teal}{Cluster $6_{14}$} is similar to the previously identified `European'
cluster, except for an important decrease in the rankings of San and Pygmies and
an increase in the rankings of Mamanwa, it's distribution resembles much that
observed at $K=12$ (cluster $5_{12}$).

\textcolor{blue!66!orange!63}{Cluster $7_{14}$} is a `South-East Asian' cluster, most predominant in
Bidayuh from Borneo. It is present at a notable level in Austronesian
populations (except those from Taiwan and the Philippines), some Austro-Asiatic
populations, JKL and Tai-Kadai.\\
Among Austronesians, it is more important in the populations of Borneo (Bidayuh
and Dayak), Java (Javanese and Sunda) and the Malaysian peninsula (Temuans,
Malays) and much less important in some non-Filipino populations of the
Philippines. Among Austro-Asiatic, it is more important in Ht'in Mal and
Palaungic and very low in Kensiu Negritos and Mlabri. Among Tai-Kadai, it is
less important in the eastern populations (Zhuang and Jiamao).

\textcolor{blue!66!orange}{Cluster $8_{14}$} is another `Austronesian' cluster, which is somewhat
complementary to the previous one. It is most important in the Philippines,
Taiwan, Sulawesi (Toraja) and Sumatra (Mentawai, Batak and Malays\footnote{The
Malay individuals were sampled in both peninsular Malaysia and Sumatra.}). It
is present at a notable level in Tai-Kadai, Chinese and Hmong-Mien. Among
Tai-Kadai, it is more important in the eastern populations, and among Chinese,
it is less important in northern populations. Cluster $8_{14}$ is also present in
other Sino-Tibetan populations, but at lower levels, and in Cambodians, Mon, Japanese,
Koreans and Melanesians.

\textcolor{orange!75!blue!63}{Cluster $9_{14}$} corresponds to the `Mlabri-specific' cluster previously
identified. It constitutes almost entirely the profile of Mlabri. It is
slightly above $9\%$ in Ht'in Mal, slightly above $7\%$ in Temuans and is
otherwise present at a low level in various populations of South-East Asia.

\textcolor{orange!75!blue}{Cluster $10_{14}$} corresponds to the `Malaysian Negrito' cluster previously
identified, but with the notable difference that it disappears from the
profile of Onge. It also decreases in Great Andamanese, the
Austronesian populations of Java, Borneo and the Malaysian peninsula, Austro-Asiatic
(except Mlabri) and JKL populations.

\textcolor{orange!66!blue}{Cluster $11_{14}$} is a `southern East Asian' cluster somewhat similar to
cluster $10_{12}$. Like cluster $10_{12}$, it has its highest proportion in Hmong,
but there are significant differences. Decreases are observed in Austro-Asiatic
populations (except Malbri and Kensiu), Austronesian populations of Java,
Borneo, and peninsular Malaysia, and JKL. It increases with respect to cluster
$10_{12}$ in Taiwanese Austonesians, Hmong, She, Chinese, Tujia and the eastern
Tai-Kadai Jiamao (more than $6.5$ percentage points), and to a lesser extent in
Koreans, Japanese, Altaic, the other Hmong-Mien, Tai-Kadai and Tibeto-Burmese
populations (except JKL), the populations of the Philippines, Sulawesi and
Mentawai, Hazara and Pahari.\refstepcounter{labs}\label{suppl:Japan-Austronesian-3}

\textcolor{blue!85!orange}{Cluster $12_{14}$} is specific to Andamanese populations and Negritos from the
Philippines. It constitutes almost entirely the profile of Onge, and
more than one third of that of Great Andamanese. It is quite important in the
profiles of Negritos from the Philippines and is notable in some populations
of India (particularly Dravidian, tribal or lower caste populations).

\textcolor{cyan}{Cluster $13_{14}$} is similar to the previously identified `Oceanian'
cluster, but almost disappears from Onge and is halved in Great Andamanese. A
significant rank decrease can be noticed in Okinawans, Srivastava and
Vaish.\refstepcounter{labs}\label{suppl:Okinawa-Andaman-2}

\textcolor{violet}{Cluster $14_{14}$} is similar to the `American' cluster previously
identified, with a strong relative decrease in Onge.

\subsection*{K = 15\refstepcounter{labs}\label{K15}\pdfbookmark[1]{K=15}{annex:K15}}

Raw results:
\texttt{Frappe\_K15.txt}\footnote{%
\url{http://dx.doi.org/10.6084/m9.figshare.117}}

Profiles of the individuals:
\texttt{Frappe\_K15.pdf}\footnote{%
\url{http://dx.doi.org/10.6084/m9.figshare.289}}

Average profiles of the populations:
\texttt{Frappe\_K15\_pops.pdf}\footnote{%
\url{http://dx.doi.org/10.6084/m9.figshare.200}}

Ranked average profiles of the populations:
\texttt{Frappe\_K15\_rankings.pdf}\footnote{%
\url{http://dx.doi.org/10.6084/m9.figshare.304}}

At $K=15$, a `Middle Eastern' cluster is present, as was the case at $K=13$.
The other clusters correspond to those present at $K=14$.

Clusters \textcolor{red}{$1_{15}$} and \textcolor{purple}{$2_{15}$} are much similar to the `general Sub-Saharan'
cluster $1_{13}$ and the `Khoisan-Pygmy' cluster $2_{13}$ respectively, except for
an important rank decrease for Onge.
%
%

\textcolor{purple!63}{Cluster $3_{15}$} is much similar to the `East African' cluster $3_{13}$ except for
an important rank decrease for Onge and Great Andamanese.

\textcolor{purple!75!lime}{Cluster $4_{15}$} is similar to cluster $4_{13}$, in as much as it constitutes about
one third of the profiles of the populations of Middle East. But there
are otherwise important differences. It decreases in many populations of
Pakistan and India, as well as in some populations of the Philippines and in
Uyghur. It is reinforced in Middle East, Italy, North Africa, Maasai and Fulani.

\textcolor{lime!75!cyan!90!black}{Cluster $5_{15}$} is similar to the `Indian' cluster previously identified.
Compared to $5_{13}$, an important decrease occurs in Great Andamanese, it
disappears from Onge, and increases in Middle East and western populations of
Pakistan. Compared to $4_{14}$, a decrease occurs for Brahui and Middle Eastern
populations and a slight increase for populations of West and North Europe.
%
%
%

\textcolor{green!75!teal}{Cluster $7_{15}$} is similar to the `European' cluster $7_{13}$. There is a decrease
in populations from Middle East, Italy, Caucasus and western Pakistan, and an
increase in Kalash.









The other clusters are mostly unchanged with respect to the case where
$K=14$, with the following correspondences:\\
\begin{tabular}{|c|c|c|c|c|c|c|c|c|c|}
\hline
Cluster                & \textcolor{olive}{$6_{15}$} & \textcolor{blue!66!orange}{$8_{15}$} & \textcolor{blue!66!orange!63}{$9_{15}$} & \textcolor{orange!75!blue}{$10_{15}$} & \textcolor{orange!75!blue!63}{$11_{15}$} & \textcolor{orange!66!blue}{$12_{15}$} & \textcolor{blue!85!orange}{$13_{15}$} & \textcolor{cyan}{$14_{15}$} & \textcolor{violet}{$15_{15}$}\\
\hline
corresponds to cluster & $5_{14}$ & $7_{14}$ & $8_{14}$ & $9_{14}$ & $10_{14}$ & $11_{14}$ & $12_{14}$ & $13_{14}$  & $14_{14}$\\
\hline
\end{tabular}

\subsection*{K = 16\refstepcounter{labs}\label{K16}\pdfbookmark[1]{K=16}{annex:K16}}

Raw results:
\texttt{Frappe\_K16.txt}\footnote{%
\url{http://dx.doi.org/10.6084/m9.figshare.118}}

Profiles of the individuals:
\texttt{Frappe\_K16.pdf}\footnote{%
\url{http://dx.doi.org/10.6084/m9.figshare.290}}

Average profiles of the populations:
\texttt{Frappe\_K16\_pops.pdf}\footnote{%
\url{http://dx.doi.org/10.6084/m9.figshare.201}}

Ranked average profiles of the populations:
\texttt{Frappe\_K16\_rankings.pdf}\footnote{%
\url{http://dx.doi.org/10.6084/m9.figshare.305}}

At $K=16$, the cluster specific to the Andamanese populations again
disappears. The `Austronesian' clusters are reorganized, with the
appearance of a cluster specific to the non-Filipino populations of the
Philippines ($10_{16}$), as was the case at $K=12$. The `American'
cluster is now separated in a `northern' cluster ($15_{16}$) and a `southern' cluster
($16_{16}$).\\
The most important changes observed in the other clusters are related
to the above-mentioned cluster appearances and disappearances: They often affect
Andamanese, Negritos from the Philippines and populations from North America.

Clusters \textcolor{red}{$1_{16}$} and \textcolor{purple!63}{$2_{16}$} correspond to clusters $1_{15}$ and $2_{15}$ respectively,
except for an important rank decrease in Mamanwa and an important rank increase in Onge.

\textcolor{purple}{Cluster $3_{16}$} corresponds to cluster $3_{15}$, except for important rank
decreases in Kalash, Pima and Mamanwa, and important rank increases in Onge,
Great Andamanese, Ayta and Ovambo.

\textcolor{purple!75!lime}{Cluster $4_{16}$} is similar to cluster $4_{15}$. A significant increase occurs in many
populations where cluster $4_{15}$ was already important (West Asia, North Africa, Europe).
An important rank decrease occurs for Mamanwa, Ati, Ayta, Pima, Ovambo, Pedi and
Great Andamanese.

\textcolor{lime!75!cyan!90!black}{Cluster $5_{16}$} is similar to cluster $5_{15}$. An increase occurs in
Andamanese, in some tribal and lower caste populations of continental India and in
some Negritos from the Philippines (Ayta, Agta and Ati). This increase is particularly
important for Onge. An important rank decrease is observed for Mamanwa and Pima.

\textcolor{teal!75!green}{Cluster $6_{16}$} is similar to cluster $7_{15}$. A decrease occurs in the
populations of Middle East, North Africa, Caucasus, Italy (more in Sardinia,
less in the north) and western Pakistan. The decreases somewhat reflect the
increases observed for the `Middle Eastern' cluster. Important rank
decreases affect Pima, the Negritos from the Philippines Mamanwa and Agta,
and some populations of Southern Africa (Herero, Tswana and San), and important
rank increases are observed for Onge and Ayta.

\textcolor{olive}{Cluster $7_{16}$} is similar to the `northern East Asian' cluster $6_{15}$.
Increases occur for Andamanese and for the Negritos from the Philippines Ayta,
Ati and Agta. This increase is particularly important for Onge. The North
American populations Pima and Maya and the North Asian population Yakut lose
more than $2$ percentage points. A decrease is also observed for Colombians,
Oroqen and Hezhen. For American populations, the decrease in the `northern
East Asian' cluster manifests itself also by an important rank decrease.
It is interesting to note that Yakut and Oroqen are the two northernmost
populations of the dataset. This variation correlation between northern East Asian
and North American populations might reflect some common ancestry, either dating back
from the colonization of America, either due to later exchanges.

\textcolor{blue!66!orange}{Cluster $8_{16}$} roughly corresponds to the `Taiwan-Philippine Austronesian'
cluster $9_{15}$. Compared to cluster $9_{15}$, an important decrease affects the
Negritos from the Philippines. A significant increase occurs in Hmong-Mien,
Tai-Kadai, southern Chinese and Taiwanese Austronesians. Cluster $8_{16}$ is
thus most important in Taiwanese populations, followed by Mentawai and the
non-Negrito populations of the Philippines.

\textcolor{blue!66!orange!75}{Cluster $9_{16}$} is similar to cluster $8_{15}$. An important rank decrease affects
Taiwanese Austronesians and Pima, and an important rank increase is observed in
Onge, Ayta and Mamanwa. In Onge, this corresponds to a significant increase in
percentage points.

Similarly to cluster $6_{12}$, \textcolor{blue!66!orange!50}{cluster $10_{16}$} is dominant in Mamanwa and
important in the other non-Filipino populations of the Philippines. However,
it has a higher level in these populations, as well as in Andamanese and in many
Austronesian and Austro-Asiatic populations. It is much lower in northern and
western European populations as well as in Kalash.

\textcolor{orange!75!blue}{Cluster $11_{16}$} corresponds to the `Mlabri-specific' cluster $10_{15}$, with a
slight increase in Andamanese and in the Austronesian populations of Taiwan,
and a slight decrease in Mamanwa and Pedi.

\textcolor{orange!75!blue!63}{Cluster $12_{16}$} corresponds to the `Malaysian Negrito-specific' cluster $11_{15}$,
with an important increase in Onge, and an important rank decrease in Mamanwa.

\textcolor{orange!66!blue}{Cluster $13_{16}$} is similar to cluster $12_{15}$, but with a decrease in some
southern East Asian populations, particularly in Hmong-Mien, Southern Chinese,
Tai-Kadai, and Taiwanese Austronesians. Among Tai-Kadai, the decrease is
stronger in eastern populations. The distribution of cluster $13_{16}$ is thus
slightly `flattened' with respect to that of cluster $12_{15}$. Important rank
decreases can be noticed for Pima, Onge and Agta.

\textcolor{cyan}{Cluster $14_{16}$} corresponds to cluster $14_{15}$, except for a strong decrease in
Mamanwa and a strong increase in Onge.

\textcolor{violet!85!orange}{Cluster $15_{16}$} is a `northern American' cluster. It constitutes
almost $75\%$ of the profile of Pima (Mexico), which is the
northernmost native American population in the dataset, and almost $30\%$ for
Maya and Colombians. It is a notable component of the profile of the
Mexicans sampled in Los Angeles. Apart from these populations, it is only
present at a low level, principally in some Indo-European and Altaic
populations, in Burusho and in Spiti.

\textcolor{violet}{Cluster $16_{16}$} is a `southern American' cluster. It constitutes almost
entirely the profiles of the Tupi-speaking Amazonian populations
(Surui and Karitiana). It is important in the other American populations and
decreases according to a south $>$ north gradient. Outside America, it is below
$5\%$ except in Yakut ($7.39\%$) and Oroqen ($5.34\%$), which are the two
northernmost populations of the dataset. This may be related to the decrease
observed for cluster $7_{16}$ with respect to cluster $6_{15}$.




%